%% file: mnras_template.tex
\documentclass[fleqn,usenatbib]{mnras} 

\usepackage{newtxtext,newtxmath}

\usepackage[T1]{fontenc}

\DeclareRobustCommand{\VAN}[3]{#2}
\let\VANthebibliography\thebibliography
\def\thebibliography{\DeclareRobustCommand{\VAN}[3]{##3}\VANthebibliography}

\usepackage{graphicx}	
\usepackage{amsmath}	
\usepackage{comment}
\usepackage{xcolor}
\usepackage{cancel}

\title[Subgrid prescriptions for star formation]{Tests of subgrid models for star formation using simulations of isolated disk galaxies}

\author[F. S. J. Nobels et al.]{
\parbox{\linewidth}{
Folkert S. J. Nobels$^{1}$\thanks{E-mail: nobels@strw.leidenuniv.nl},
Joop Schaye$^{1}$,
Matthieu Schaller$^{1,2}$,
Sylvia Ploeckinger$^{2,3}$, \newline
Evgenii Chaikin$^{1}$,
and Alexander J. Richings$^{4,5,6}$}
\vspace{1mm}\\
$^{1}$Leiden Observatory, Leiden University, PO Box 9513, NL-2300 RA Leiden, the Netherlands\\
$^{2}$Lorentz Institute for Theoretical Physics, Leiden University, PO Box 9506, NL-2300 RA Leiden, The Netherlands \\
$^{3}$Department of Astrophysics, University of Vienna, Türkenschanzstrasse 17, 1180 Vienna, Austria \\
$^{4}$Institute for Computational Cosmology, Department of Physics, Durham University, South Road, Durham DH1 3LE, UK \\
$^{5}$E. A. Milne Centre for Astrophysics, University of Hull, Cottingham Road, Hull, HU6 7RX, UK \\
$^{6}$DAIM, University of Hull, Cottingham Road, Hull, HU6 7RX, UK
}

\date{Accepted XXX. Received YYY; in original form ZZZ}

\pubyear{2021}

\begin{document}
\label{firstpage}
\pagerange{\pageref{firstpage}--\pageref{lastpage}}
\maketitle

\begin{abstract}
\noindent We use smoothed-particle hydrodynamics simulations of isolated Milky Way-mass disk galaxies that include cold, interstellar gas to test subgrid prescriptions for star formation (SF). Our fiducial model combines a Schmidt law with a gravitational instability criterion, but we also test density thresholds and temperature ceilings. While SF histories are insensitive to the prescription for SF, the Kennicutt-Schmidt (KS) relations between SF rate and gas surface density can discriminate between models. We show that our fiducial model, with an SF efficiency per free-fall time of 1 per cent, agrees with spatially-resolved and azimuthally-averaged observed KS relations for neutral, atomic and molecular gas. Density thresholds do not perform as well. While temperature ceilings selecting cold, molecular gas can match the data for galaxies with solar metallicity, they are unsuitable for very low-metallicity gas and hence for cosmological simulations. We argue that SF criteria should be applied at the resolution limit rather than at a fixed physical scale, which means that we should aim for numerical convergence of observables rather than of the properties of gas labelled as star-forming. Our fiducial model yields good convergence when the mass resolution is varied by nearly 4 orders of magnitude, with the exception of the spatially-resolved molecular KS relation at low surface densities. For the gravitational instability criterion, we quantify the impact on the KS relations of gravitational softening, the SF efficiency, and the strength of supernova feedback, as well as of observable parameters such as the inclusion of ionized gas, the averaging scale, and the metallicity.

\end{abstract}




\section{Introduction}
\input{intro.tex}

\input{subgrid_model.tex}

\section{Results} \label{sec:results}
In \S~\ref{subsec:fid} we will discuss the results of the fiducial model followed, in \S~\ref{subsec:subgrid_variations}, by variations in the subgrid prescriptions for SF and stellar feedback in order to understand the impact that different aspects of the galaxy formation model have. 

\input{KS_relation}
\input{subgrid_variations}

\section{Discussion} \label{sec:discussion}
\input{softening.tex}

\input{resolution.tex}

\input{conclusion.tex}

\section*{Acknowledgements}
We would like to thank Sara Ellison, Ismael Pessa, Miguel Querejeta and Andreas Schruba for kindly providing us their KS relation data points for comparing with our simulations. 
The research in this paper made use of the \texttt{SWIFT} open-source simulation code (\url{http://www.swiftsim.com}, \citealt{schaller2018}) version 0.9.0.
This work used swiftsimio \citep{borrow2020b} for reading and visualising the data. SPH projections use the subsampled projection backends \citep{borrow2021}. This work used the DiRAC$@$Durham facility managed by the Institute for Computational Cosmology on behalf of the STFC DiRAC HPC Facility (www.dirac.ac.uk). The equipment was funded by BEIS capital funding via STFC capital grants ST/K00042X/1, ST/P002293/1, ST/R002371/1 and ST/S002502/1, Durham University and STFC operations grant ST/R000832/1. DiRAC is part of the National e-Infrastructure.
EC is supported by the funding from the European Union’s Horizon 2020 research and innovation programme under the Marie Skłodowska-Curie grant agreement No 860744 (BiD4BESt).

\section*{Data Availability}

The initial conditions of the simulations presented in this paper are publicly available in \texttt{SWIFT}. \texttt{SWIFT} is publicly available and can be found at \url{www.swiftsim.com}. The other data underlying this article will be shared on a reasonable request to the corresponding author.




\bibliographystyle{mnras}
\bibliography{example} 




\appendix

\section{Metallicity variations temperature criterion}
\label{app:temp-metals}
\begin{figure}
	\includegraphics[width=\columnwidth]{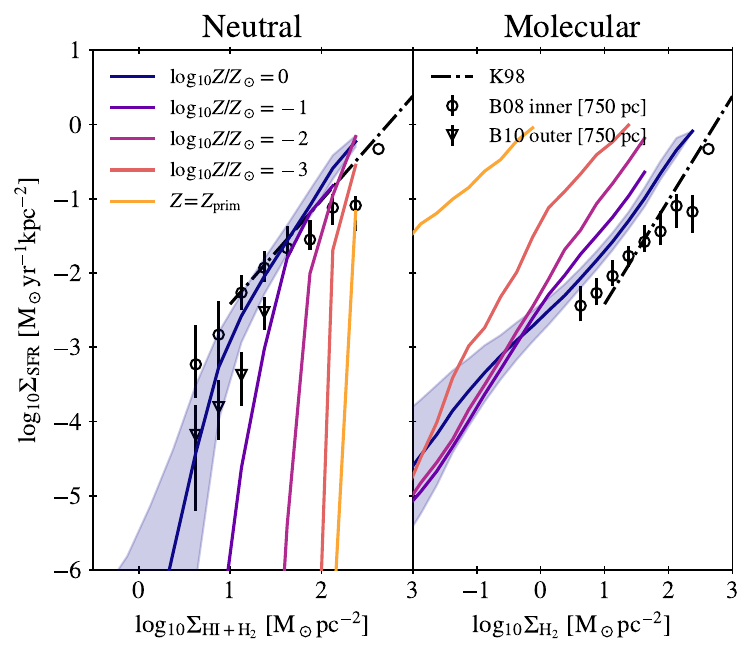} \vspace{-5mm}
    \caption{Comparison of the resolved (750 pc) neutral and molecular KS relation at different metallicities (different colours) for the temperature criterion. The shaded region shows the 16\textsuperscript{th} and 84\textsuperscript{th} percentile scatter for the solar metallicity simulation. We show the disk averaged KS relation from \citetalias{kennicutt1998}. The neutral KS relation changes almost two dex in the position of the break while the molecular KS relation has a change in normalization of more than two orders of magnitude compared to solar metallicity.}
    \label{fig:KS-importance-Z-temp}
\end{figure}

Fig. \ref{fig:KS-importance-Z-temp} shows the metallicity dependence of the KS relation for a temperature criterion ($T<10^3~\rm K$). The dependence on metallicity is slightly stronger than for a gravitational instability criterion.


\bsp	
\label{lastpage}
\end{document}

%% file: intro.tex
Cosmological hydrodynamical simulations of representative volumes typically only resolve scales of $\sim 0.1 - 1$ kpc and masses of $\sim 10^4 - 10^6~\rm M_\odot$ \citep[e.g.][]{schaye2015,pillepich2018,nelson2019,dave2019,dubois2021,feldmann2022}.  Because much of the physics of star formation (SF) remains unresolved, idealised SF laws are used to determine the SF rates (SFRs) of individual resolution elements. Generally, a so-called \citet{schmidt1959} law is adopted, which assumes that gas collapses on the density-dependent free-fall time, $t_{\rm ff}=\sqrt{3\pi/32G\rho}$, and is converted into stars with a specified (fractional) SF efficiency (SFE) per free-fall time, $\varepsilon$,
\begin{equation}
    \dot{\rho}_\star = \varepsilon \frac{\rho}{t_{\rm ff}}. \label{eq:schmidt-law}
\end{equation}
The SFE $\varepsilon$ is a free parameter, typically set between 1 per cent \citep[e.g.][]{semenov2017} and 100 per cent \citep[e.g.][]{hopkins2018}.

Alternative approaches can however also be found in the literature. For example, the \citet{schaye2008} subgrid model uses the gas pressure to set the SFR of a resolution element. For simulations that do not resolve the multiphase interstellar medium (ISM) advantages of this approach are that, unlike the density, the pressure does not change dramatically between simulations that do and do not resolve a cold (molecular) gas phase, and that for a self-gravitating disk the pressure is closely related to the gas surface density that appears in the observed kpc-scale \citet{kennicutt1998} SF surface density law. However, these features are arguably undesirable for high-resolution models that capture the physics resulting in a multiphase ISM and which can be tested by comparing observed and predicted coarse-grained SFEs. Other subgrid prescriptions use physically-motivated SFE models that depend not only on the density, but also on the velocity dispersion \citep[e.g.][]{semenov2016,kretschmer2020}.

Applying an SF law to all gas may be unrealistic given that stars are observed to form in cold and dense gas. Moreover, we do not wish to impose subgrid models on resolved scales. Therefore, a criterion needs to be applied to determine which gas in the simulation is eligible for SF. Typically, the selected gas satisfies one (or multiple) of the following criteria: the physical gas density exceeds a hydrogen number density threshold $n_{\rm H, crit}$ that is either constant \citep[e.g.][]{springel2003} or depends on metallicity \citep[][]{schaye2004}, is colder than a critical temperature $T_{\rm crit}$ \citep[e.g.][]{wang2015,revaz2018}, is \citet{jeans1902} unstable \citep[e.g.][]{stinson2006,hopkins2018}, is in a converging flow \citep[e.g.][]{stinson2006,hopkins2022} and/or is gravitationally bound \citep[e.g.][]{hopkins2014}. Additionally, cosmological simulations often include a criterion that only allows gas to be star-forming if it is $\sim 10^2$ times denser than the cosmic mean \citep[e.g.][]{schaye2015}. This prevents SF in the gas outside of haloes at very high redshifts where the critical physical density threshold becomes comparable to the cosmic mean.

Observationally, galaxy-averaged SFR surface density relations were shown by \citet{kennicutt1989,kennicutt1998} to follow
\begin{align}
    \Sigma_{\rm SFR} &= A \Sigma_{\ion{H}{I} + \rm H_2}^N,
\end{align}
where $\Sigma_{\rm SFR}$ is the SFR per unit area, $\Sigma_{\ion{H}{I} + \rm H_2}$ is the neutral hydrogen surface density, $A$ is the normalisation and $N\approx 1.4$ is the slope of the power law. \citet{kennicutt2007} showed that spatially-resolved kpc-sized regions in individual galaxies follow the same surface density law, which is often referred to as the Kennicutt-Schmidt (KS) relation. Over the last decade, observational surveys have measured azimuthally averaged and spatially resolved scaling relations for large collections of galaxies on spatial scales of $\sim 750~\rm pc$, thus providing valuable constraints on models of galaxy and SF \citep[e.g.][]{sanchez2012,bundy2015,fogarty2015,leroy2021,lin2019}. These include atomic KS relations, $\Sigma_{\rm SFR}=A\Sigma_{\ion{H}{I}}^N$ \citep[e.g.][]{bigiel2008,bigiel2010,Schruba2011} and molecular KS relations, $\Sigma_{\rm SFR}=A\Sigma_{\rm H_2}^N$ \citep[e.g.][]{bigiel2008,bigiel2010,onodera2010,Schruba2011,bolatto2017,lin2019, ellison2020,pessa2021, querejeta2021,abdurrouf2022}.
The observations indicate that the azimuthally-averaged SFR in disk galaxies quickly decreases beyond the optical radius \citep[e.g.][]{kennicutt1989,martin2001,koopmann2004,bigiel2010}, while the gas density does not decrease as fast because the \ion{H}{I} disk typically extends far beyond the optical disk \citep[e.g.][]{bosma1981,broeils1994,reeves2015}. As a result of this sudden drop in SF a `break' is seen in the azimuthally-averaged and spatially-resolved total gas KS relation around surface densities of $\Sigma_{\rm \ion{H}{I}+H_2}=10 ~\rm M_\odot ~\rm pc^{-2}$ \citep[e.g.][]{martin2001,bigiel2008,bigiel2010,dessauges-zavadsky2014}. The presence of a break in the KS relation is one of the strongest indications that simulations require an SF criterion.

CO observations of nearby galaxies show that on scales of $\sim 10^2~$pc molecular gas with a velocity dispersion around $14~\rm km ~\rm s^{-1}$ has almost a factor three longer molecular gas depletion time\footnote{We define the depletion time as $t_{{\rm dep},i}=\Sigma_i/\Sigma_{\rm SFR}$, where $i$ is the gas component under consideration: neutral hydrogen, molecular hydrogen or atomic hydrogen.} than gas with velocity dispersions of $12~\rm km ~\rm s^{-1}$ \citep{leroy2017}. Similarly, spatially-resolved ($10^2~\rm pc$) observations that target denser gas (traced by HCN and/or HCO$^{+}$) show that a higher velocity dispersion correlates with a longer molecular gas depletion time scale \citep[e.g.][]{murphy2015,viaene2018,querejeta2019}. Furthermore, observations of individual molecular clouds show that the observed depletion times are spread over around two orders of magnitude \citep[e.g.][]{heiderman2010, murray2011, evans2014, lee2016, vutisalchavakul2016, ochsendorf2017, pokhrel2021} and that more massive giant molecular clouds (GMCs), which according to \citeauthor{larson1981}'s \citeyearpar{larson1981} empirical laws, have higher velocity dispersions, have longer depletion time scales \citep[e.g.][]{lee2016,ochsendorf2017}. These observations hence indicate that on scales of $100~\rm pc$, gas with a higher velocity dispersion is less likely to form stars. This suggests that an SF criterion based on the velocity dispersion or gravitational instability of gas might be in better agreement with observations than e.g.\ a constant density threshold. 

Numerical simulations have been performed to investigate the diversity of the depletion time in GMCs. They found that the gas depletion time in GMCs varies over more than one order of magnitude \citep[e.g.][]{padoan2012,grisdale2019,grudic2019}, that the average GMC depletion time scale is consistent with an SFE of 1 per cent \citep[e.g.][]{grisdale2019,grudic2019} (though some theoretical works predict higher SFEs e.g. \citealt[][]{raskutti2016}). In addition, they predict that the diversity of the SFE can be caused by several factors such as the gravitational boundedness of the GMCs \citep{padoan2012}, the large diversity in the cloud properties \citep{grisdale2019}, and the mass of the GMC \citep{grisdale2021}. Simulations of galaxies have been performed using gas depletion times that depend on the boundedness of clouds \citep{semenov2016, gensior2020, kretschmer2020}, or simply using an SF criterion based on the boundedness of the cloud \citep[e.g.][]{hopkins2013,hopkins2014,hopkins2018,hopkins2022,semenov2017, semenov2018,semenov2019}. Not much attention has however been given to comparing them to KS relations split by hydrogen species or different spatially-averaging approaches.

To test subgrid models for SF, simulations of idealised, isolated galaxies offer some advantages over full, cosmological simulations. They offer more control, are less computationally expensive, and do not suffer from the chaotic behavior due to e.g.\ slight changes in the merger history that complicated comparisons of individual objects in different cosmological simulations using identical initial conditions \citep{genel2019,keller2019,borrow2022}. 
These features make such simulations well suited for systematic explorations of parameter space and convergence tests. 

In this work, we use hydrodynamical simulations of isolated disk galaxies that include a cold, molecular gas phase to systematically investigate several criteria for SF. For our fiducial subgrid model, which consists of a Schmidt law combined with a gravitational instability criterion, we compare with observations of both azimuthally averaged and spatially resolved KS laws for neutral, atomic and molecular gas. We systematically investigate the effect of the choice of SF criterion, the parameter values, and the numerical resolution.

The remainder of this paper is structured as follows. In Section \ref{sec:sims} we describe our subgrid model for ingredients other than the SF criterion and our disk galaxy set-up. This is followed by an overview of the different criteria for SF that we compare: a density threshold, a temperature ceiling, and a gravitational instability criterion. In Section \ref{sec:results} we show our results and in Section \ref{sec:discussion} we discuss the effect of gravitational softening and compare our results with previous work. In Section \ref{sec:conclusion} we summarise our conclusions.

%% file: subgrid_model.tex
\section{Simulations} \label{sec:sims}

We perform hydrodynamical simulations of isolated disk galaxies using the publically available code \texttt{SWIFT} \citep{schaller2016,schaller2018,schaller2023}. We use a fast multipole method \citep[][]{greengard1987} as the gravity solver together with a static external \citet{hernquist1990} potential. For the hydrodynamics, we use the SPHENIX scheme and parameter values of \citet{borrow2020}. SPHENIX is a density-energy smoothed-particle hydrodynamics (SPH) scheme designed to capture shocks and contact discontinuities with artificial viscosity following \citet{cullen2010} and artificial conduction following \citet{price2012}. We use a quartic spline for the SPH kernel with the mean weighted number of neighbours given by $\langle N_{\rm ngb} \rangle=65$. The time steps of particles are limited by the local acceleration ($\Delta t \propto 1/\sqrt{|\mathbf{a}|}$) and limited to 1 per cent of the circular orbital period at the particle's position \citep[similar to what was done in][]{nobels2022}. Additionally, for gas, the time steps are limited by the Courant-Friedrichs-Lewy (CFL) condition ($C_{\rm CFL}=0.2$) and the \citet{durier2012} time step limiter is used to prevent vastly different time steps between neighbouring gas particles. We use a fiducial baryonic resolution of $10^5~\rm M_\odot$, with a corresponding gravitational \citeauthor{plummer1911}-equivalent softening length of $200~\rm pc$. Lastly, we limit the smoothing length to a minimum of $1.55 \times 2~\rm pc = 3.1~\rm pc$ to prevent artifical collapse (see Ploeckinger et al. in prep.) and a maximum of $10~\rm kpc$.
\nocite{courant1928}

\subsection{Radiative cooling}

Radiative cooling and heating rates are calculated using the non-equilibrium chemistry network \texttt{Chimes} \citep{richings2014a, richings2014b} for hydrogen and helium species and free electrons. The radiation field consists of a redshift-dependent metagalactic UV background \citep{faucher-giguere2020} and an interstellar radiation field whose intensity increases with the Jeans column density to the power of 1.4, following the KS relation (see \citealt{ploeckinger2020}, Ploeckinger et al, in prep. for details). Radiation is shielded by a Jeans column density that accounts for both a thermal and a constant turbulent (velocity dispersion of $6~\mathrm{km~s}^{-1}$) pressure component.
The cosmic ray rate follows the same scaling as the interstellar radiation field for low column densities but saturates at a cosmic ray rate of $2\times10^{-16}~\mathrm{s}^{-1}$ for $N_{\mathrm{H}} \ge 10^{21}\,\mathrm{cm}^{-2}$ \citep{indriolo2015}. In order to account for unresolved clumping, the reactions on the surface of dust grains, such as the formation of $\rm H_2$, are boosted above $0.1~\rm cm^{-3}$ as $B_{\rm dust} = 10^{1/3} ( n_{\rm H} / 1 \,\rm cm^{-3} )^{1/3}$, the boost factor saturates at $B_{\rm dust} = 10$ for densities $n_{\rm H}\geq 100~\rm cm^{-3}$.  

The cooling and heating rates of C, N, O, Ne, Mg, Si, S, Ca, and Fe are pre-tabulated using the same radiation field and shielding length assumptions as for the non-equilibrium rates of hydrogen and helium. Their ion fractions are calculated using \texttt{Chimes} assuming chemical equilibrium and the rates are calculated based on the individual species fractions. A constant dust-to-metal ratio of $5.6\times 10^{-3} Z/Z_\odot$ is assumed for neutral gas with a cut-off towards higher temperatures ($T > 10^5\,\mathrm{K}$) for which thermal sputtering destroys dust grains on short timescales. The complete set of cooling tables will be made public with a forthcoming publication (Ploeckinger et al. in prep.).

\subsection{Star formation rates}\label{sec:SFRs}
When gas is selected to be star-forming (see \S \ref{sec:starformation} for a description of the different SF criteria that we employ), it is assigned an SFR following a \citet{schmidt1959} law (equation~\ref{eq:schmidt-law}). We use a constant SFE $\varepsilon=0.01$, which is motivated by observations of GMCs in the Milky Way \citep[e.g.][]{vutisalchavakul2016,lee2016,pokhrel2021,zipeng2021}, the Large Magellanic Cloud \citep[e.g.][]{ochsendorf2017} and, other nearby galaxies \citep[e.g.][]{utomo2018}. In \S~\ref{subsubsec:SFE} we investigate the impact that the SFE has on the observed KS relations.  

Because the gas consumption time scale $\rho/\dot{\rho}_\star = t_{\rm ff}/\varepsilon$ is much longer than the typical time step size, the Schmidt law needs to be implemented stochastically and the probability of a gas particle converting to a star particle during a time step $\Delta t$ is:
\begin{align}
    \text{Prob} = {\rm min} \left( \frac{\dot{\rho}_\star \Delta t}{\rho}, 1 \right) = {\rm min} \left( \frac{\varepsilon \Delta t}{t_{\rm ff}}, 1 \right).
\end{align}

\subsection{Stellar feedback}

After gas particles have been stochastically converted into stellar particles with mass $m_\star$, they are assumed to represent simple stellar populations that follow a \citet{chabrier2003} initial mass function (IMF) with a zero age main sequence mass range of $0.1-100~\rm M_\odot$. We turn off chemical enrichment to ensure that the simulations have a constant metallicity similar to the observations.

\subsubsection{Early stellar feedback}
We account for three pre-supernova feedback processes: \ion{H}{II} regions, stellar winds, and radiation pressure. The implementation of early stellar feedback, whose effects are modest compared to those of supernova feedback, will be described in detail by Ploeckinger et al. (in prep.). We will only give a short explanation of each process below.

The total mass of the \ion{H}{II} regions is based on the expected size of the \citet{stromgren1939} sphere given the local density and the average flux of hydrogen ionizing photons per unit stellar mass during the time step, which is obtained from the Binary Population and Spectral Synthesis (BPASS) stellar evolution and spectral synthesis models \citep{eldridge2017,stanway2018}. Gas in \ion{H}{II} regions is assumed to be ionized, non star-forming, and its temperature is limited to $\geq 10^4~\rm K$. The radiation pressure is computed using the BPASS photon energy spectrum and the wavelength-dependent optical depth $\tau(\lambda)$ implied by the shielding column used for the cooling tables of \citet{ploeckinger2020} and the tables used here, which scales with the local Jeans length. The momentum from stellar winds is taken from the BPASS tables. The momentum injected during a time step by radiation pressure and stellar winds from a star particle is injected by  stochastically kicking neighbouring gas particles with a velocity $v_{\rm kick}=50~\rm km~\rm s^{-1}$ radially away from the stellar particle. 

\subsubsection{Supernova feedback }
We assume that stars with initial mass $> 8~\rm M_\odot$ produce core-collapse supernovae (CC SNe), which for our IMF corresponds to $1.18\times 10^{-2}~\rm M_\odot^{-1}$ CC SN per unit stellar mass formed. Each CC SN provides a total energy of $E_{\rm CC SN} = 2.0 \times 10^{51}~\rm erg$. We choose the same value as we used in \citet{chaikin2022b}. This value is slightly higher than the canonical $10^{51}~\rm erg$, but may be viewed as accounting for the $\sim10$ times more energetic hypernovae, stars more massive than 100~M$_\odot$, and/or simply as compensation for some degree of numerical overcooling.  We use metallicity-dependent stellar lifetime tables to calculate the number of CC SN that explode every time step \citep[based on][]{portinari1998}. Following \citet{chaikin2022b}, star particles inject 90 per cent of their CC SN energy in thermal form and the remaining 10 per cent as kinetic energy. The thermal part is injected using the stochastic method of \citet{dallavecchia2012} using the statistically isotropic selection of gas particles of \citet{chaikin2022}. The temperature of the gas receiving thermal CC SN feedback is increased by $\Delta T=10^{7.5}~\rm K$ to suppress numerical overcooling. The remaining 10 per cent of the energy is injected kinetically using the energy, momentum and angular momentum conserving method detailed in \citet{chaikin2022b}. Pairs of particles are selected isotropically and kicked in opposite directions with a desired velocity of $\Delta v_{\rm CC SN}=50~\rm km ~\rm s^{-1}$ (the actual kick velocity depends on the relative velocity of the star and gas particles). The high-energy thermal events drive galactic winds and can be thought of as representing superbubbles due to clustered CC SN events, while the low-energy kinetic events drive turbulence and can be viewed as representing the momentum-driven phase of isolated CC SN. For more details and discussion we refer to \citet{chaikin2022b}.

For type-Ia supernova (SNIa) feedback, the exact delay of SNIa cannot be predicted because it depends on poorly constrained parameters (i.e.\ the binary fraction and separation). Therefore, we use a statistical approach that samples the SNIa rate from a delay time distribution (DTD) as
\begin{align}
 \text{DTD}(t) &= \frac{\nu}{\tau}e^{-(t-t_\text{delay})/\tau}
  \Theta(t-t_\text{delay}),
\end{align}
where $\nu=2.0 \times 10^{-3} ~\rm M_\odot^{-1}$ is the number of SNIa per unit formed stellar mass \citep{maoz2012}, $\tau = 2~\rm Gyr$ \citep[based on Fig 2 of][]{maoz2010}, $\Theta(x)$ is the Heaviside step function and $t_\text{delay}=40~\rm Myr$ corresponds to the maximum lifetime of a star that explodes as a CC SN \citep{portinari1998}. SNIa energy is injected 100 per cent thermally and isotropically as described above for CC SN, but using $10^{51}\,$erg per SNIa. 

\input{initial_conditions.tex}

\section{Star formation criteria}
\label{sec:starformation}
In this work we investigate the impact of different subgrid prescriptions for SF, in particular different criteria for selecting gas particles that are eligible for conversion into stellar particles. The calculation of the SFRs of star-forming gas particles was described in \S \ref{sec:SFRs}. This section describes how the star-forming gas is selected. We will test three different criteria: a density threshold (\S\ref{sec:density_threshold}), a temperature ceiling (\S\ref{sec:temperature_ceiling}) and a gravitational instability criterion (\S\ref{subsub:instability}). 

\begin{figure}
    \centering
    \includegraphics[width=\columnwidth]{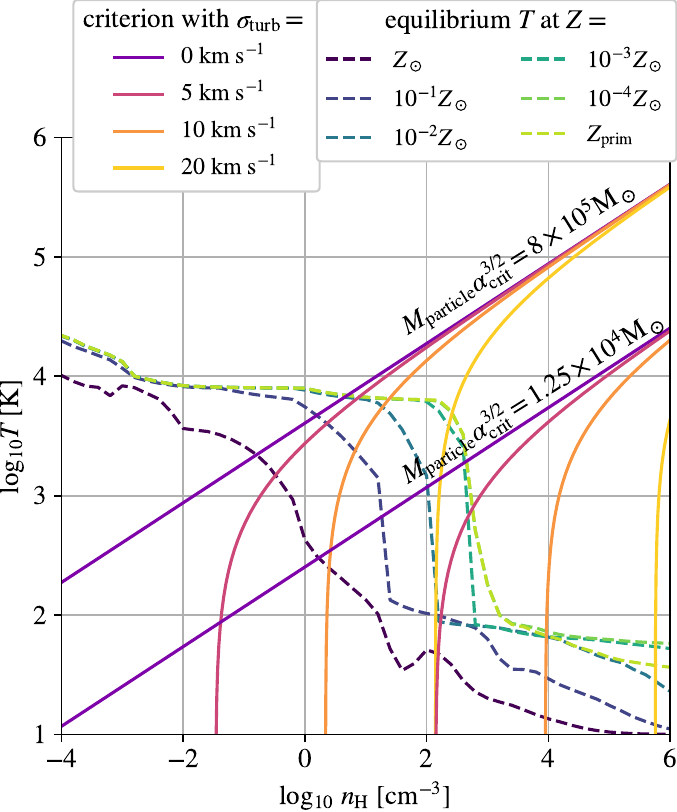} \vspace{-5mm}
    \caption{The gravitational instability criterion in temperature-density space for different velocity dispersions (solid lines, different colours; gas at higher densities or lower temperatures is unstable) and different resolutions and/or normalisations (two sets of lines labelled with different values of $M_{\rm particle} \alpha_{\rm crit}^{3/2}$). For comparion we also plot the thermal equilibrium temperature curves for different metallicities (dotted lines, different colours) for the  Ploeckinger et al. (in prep.) cooling tables. The gray lines indicate different density thresholds and different temperature ceilings. Gas with a higher velocity dispersion requires higher densities to be star-forming. The gas selected by the gravitational instability criterion shifts to higher (lower) densities for higher (lower) numerical resolutions, i.e.\ lower (higher) values of $M_\text{particle}$, if the instability criterion (i.e., the value of $\alpha_{\rm crit}$) is kept constant. Therefore, the gas phase that is selected to be star-forming depends on the numerical resolution, unless we scale $\alpha_{\rm crit}\propto M_\text{particle}^{-2/3}$.
    }
    \label{fig:sfcriteria}
\end{figure}

\subsection{Density threshold} \label{sec:density_threshold}
The simplest and most widely used SF criterion is a density threshold,
\begin{align}
    n_{\rm H} > n_{\rm H, \rm crit},
\end{align}
where $n_{\rm H, \rm crit}$ is the total hydrogen number density\footnote{We use a fixed primordial hydrogen mass fraction $X=0.756$ to convert the gas mass density into the total hydrogen number density.} above which the gas is star-forming. 

It is instructive to see what gas phases are selected by criteria with different density thresholds.
The dotted curves in Fig.~\ref{fig:sfcriteria} show the equilibrium temperature\footnote{The equilibrium temperature is the temperature at which the radiative heating rate is equal to the radiative cooling rate.} as a function of density for different metallicities. We see that for $Z=Z_\odot$ the transition to the cold phase ($T\ll 10^3~\rm K$) occurs at $n_{\rm H}\sim 10^{-0.5}~\rm cm^{-3}$ while for very low metallicities (e.g. $Z=10^{-4} Z_\odot$) the equilibrium temperature remains above $10^3~\rm K$ for densities $n_{\rm H} \lesssim 10^3~\rm cm^{-3}$. This means that for different metallicities the selected gas will correspond to different phases, i.e.\ at low metallicities the star-forming gas is warm, while at higher metallicities it is cold. 

Since observations indicate that SF is associated with cold gas, a constant density threshold has the disadvantage that, depending on the metallicity, the selected gas can be in the warm phase unless the threshold is chosen to be very high. A fixed density threshold is common in the literature \citep[e.g.][and models based on this]{springel2003}, while other models use a metallicity-dependent density threshold designed to track the transition from the warm to the cold interstellar gas-phase (e.g.\ \citealt[][]{schaye2004, agertz2013} and models based on these such as EAGLE; \citealt[][]{schaye2015}).

\subsection{Temperature ceiling} \label{sec:temperature_ceiling}
To restrict the formation of stars to colder gas that resembles the environment where stars are observed to form (e.g.\ molecular clouds), a temperature criterion is a natural choice,
\begin{equation}
    T < T_{\rm crit}.
\end{equation}
The parameter $T_{\rm crit}$ defines a temperature ceiling, which is usually set to be low enough to only select gas that is expected to have conditions suitable for SF. Looking again at the equilibrium temperature curves in Fig.~\ref{fig:sfcriteria}, we see that for very low metallicities a reasonable value like $T_{\rm crit}=10^2~\rm K$ will only select gas with very high densities, $n_{\rm H} \gtrsim 10^4~\rm cm^{-3}$. If such high densities are not resolved, then the SFR will be underestimated. Cosmological simulations with limited resolution that start with primordial abundances may never form any stars at all. Combining density and temperature criteria, i.e.\ $n_{\rm H} > n_{\rm H, \rm crit}$ or $T < T_{\rm crit}$, can help alleviate this problem. We will, however, consider a different solution.

\subsection{Gravitational instability criterion} \label{subsub:instability}
Another option for the SF criterion is to select gas that is gravitationally unstable, i.e. that satisfies the \citet{jeans1902} instability criterion $t_{\rm ff} < t_{\rm cr}$, where $t_{\rm cr}$ is the sound and turbulence crossing time which for a cloud of size $h$ is\footnote{Note that it is possible to include additional terms, for example when MHD is used (e.g. $ t_{\rm cr} = h/\sqrt{\sigma_{\rm th}^2 + \sigma_{\rm Alfven}^2 + \sigma_{\rm turb}^2}$), where $\sigma_{\rm Alfven}$ is the velocity dispersion of ions.}
\begin{align}
  t_{\rm cr} &= \frac{h}{\sqrt{\sigma_{\rm th}^2 + \sigma_{\rm turb}^2}}, \label{eq:soundcross}
\end{align}
where the thermal dispersion $\sigma_{\rm th}$ is given by:
\begin{align}
    \sigma_{\rm th} &= \sqrt{3 \frac{P}{\rho}} = \sqrt{\frac{3 k_{\rm B} T}{\mu m_{\rm H}}} = 13.8{~\rm km ~ s^{-1}} \left( \frac{T}{10^4~\rm K} \right)^{1/2},
\end{align}
where we set $\mu=1.3$ which is appropriate for neutral atomic gas, and $\sigma_{\rm turb}$ is the turbulent velocity dispersion. In our SPH implementation, the turbulent velocity dispersion is calculated for each particle $i$ and its neighbours $j$ as:
\begin{equation}
    \sigma_{{\rm turb}, i}^2 = \frac{1}{\rho_i} \sum\limits_{j=1}^N M_{{\rm particle},j} W(r_{ij}, s_i) |\mathbf{v}_i - \mathbf{v}_j|^2, \label{eq:veldis}
\end{equation}
where $M_{\rm particle}$ is the particle mass, $W$ is the kernel weight function, $s$ is the smoothing length, $\mathbf{v}_i$ and $\mathbf{v}_j$ are the particle velocities, and $r_{ij}$ is the particle separation. We find that this gives identical results to the velocity dispersion calculations of \citet{hopkins2013,hopkins2018} who compute either a full tensor or both $\nabla \times \mathbf{v}$ and $\nabla \cdot \mathbf{v}$. An advantage of using equation (\ref{eq:veldis}) is that only a single SPH variable is required in contrast with 4 or 9 additional SPH variables\footnote{Six additional variables when symmetry arguments are used.}.

A cloud is gravitationally unstable when its free-fall time is smaller than the sound and turbulence crossing time. We calculate the free-fall time from first principles for consistency with \S \ref{subsec:softening} where we will investigate the impact of including the gravitational softening length $\epsilon$ in the derivation. The free-fall time is given by
\begin{align}
 t_{\rm ff} &= \int\limits_h^0 \frac{{\rm d}t}{{\rm d}r} {\rm d}r.
\end{align}
For an initially static, spherical cloud of mass $m$ and radius $h$, energy conservation implies that the velocity of the outer shell is given by
\begin{align}
  v(r) &= -\left( \frac{2G m}{r} - \frac{2G m}{h}\right)^{1/2}. \label{eq:velocity}
\end{align}
This means that the free-fall time is given by
\begin{align}
    t_{\rm ff} &= \frac{1}{\sqrt{2Gm}} \int\limits_0^h \left( \frac{1}{r} - \frac{1}{h} \right)^{-1/2} {~\rm d} r, \\
    &= \sqrt{\frac{3}{8\pi G\rho}} \int\limits_0^1 \sqrt{\frac{x}{1-x}}{~\rm d} x, \\
    &= \sqrt{\frac{3\pi}{32 G\rho}} =45 ~{\rm Myr} \left( \frac{n_{\rm H}}{1 ~\rm cm^{-3}} \right)^{-1/2}. \label{eq:freefall}
\end{align}
Equating $t_\text{cr}$ (equation~\ref{eq:soundcross}) and $t_\text{ff}$ (equation~\ref{eq:freefall}) gives the generalized \citet{jeans1902} length:
\begin{align}
    \lambda_{\rm J} &= \sqrt{\frac{3\pi(\sigma_{\rm th}^2 + \sigma_{\rm turb}^2)}{32 G \rho}}, \\
    &= 0.46 {~\rm kpc} \left( \frac{\sqrt{\sigma_{\rm th}^2 + \sigma_{\rm turb}^2}}{10~{\rm km ~\rm s^{-1}}} \right) \left( \frac{n_{\rm H}}{1 ~{\rm cm^{-3}}} \right)^{-1/2},
\end{align}
and a corresponding \citet{jeans1902} mass:
\begin{align}
    M_{\rm J} &= \frac{4\pi}{3}\rho \lambda_{\rm J}^3 = \frac{4\pi}{3} \left( \frac{3\pi}{32}\right)^{3/2} \left( \frac{\sigma_{\rm th}^2 + \sigma_{\rm turb}^2}{G} \right)^{3/2} \rho^{-1/2}, \label{eq:jeansmass} \\
    &= 1.3 \times 10^7 {~\rm M_\odot} \left( \frac{\sqrt{\sigma_{\rm th}^2 + \sigma_{\rm turb}^2}}{10~{\rm km ~\rm s^{-1}}} \right)^3 \left( \frac{n_{\rm H}}{1 ~{\rm cm^{-3}}} \right)^{-1/2}.
\end{align}
We can construct an SF criterion by demanding gravitational instability at the mass resolution limit, i.e., by requiring the \citet{jeans1902} mass to be smaller than the expected mass within the SPH kernel
\begin{align}
    M_{\rm J} < M_{\rm res} = \langle N_{\rm ngb} \rangle M_{\rm particle}. \label{eq:jeansmass-crit}
\end{align}
Combining equations (\ref{eq:jeansmass}) and (\ref{eq:jeansmass-crit}), we get
\begin{align}
    \alpha \equiv \frac{\sigma_{\rm th}^2 + \sigma_{\rm turb}^2}{G \langle N_{\rm ngb} \rangle^{2/3} M_{\rm particle}^{2/3} \rho^{1/3}} < \alpha_{\rm crit}, \label{eq:instability-crit}
\end{align}
where $\alpha_{\rm crit}$ is a constant of order unity and the gas is considered unstable if $
\alpha < \alpha_\text{crit}$. Using equation (\ref{eq:jeansmass}) gives $\alpha_{\rm crit} = 32/3 \pi  (3/4\pi)^{2/3}\approx 1.3$, but the appropriate value depends on the assumed geometry. Using a smaller values of $\alpha_{\rm crit}$ implies that star forming gas must be more strongly gravitationally bound. Our fiducial value is $\alpha_{\rm crit}=1$. 

Similar criteria have been used before in hydrodynamical simulations, for example by  \citet{hopkins2014, hopkins2018} and \citet{semenov2016}, and are often called virial criteria. However, our calculation of the gravitational instability criterion is not identical to the ones in those papers. The main difference is that they used the sound speed instead of the thermal dispersion and that they adopted another normalisation. The difference in normalisation is because a gravitational instability criterion requires gas to satisfy $E_{\rm grav} + E_{\rm kin} < 0$ while a virial criterion requires gas to satisfy $E_{\rm grav} + 2 E_{\rm kin} < 0$.

Based on equation (\ref{eq:instability-crit}), it is clear that for a fixed $\alpha_{\rm crit}$ the instability criterion selects different gas for different resolutions $M_{\rm particle}$. We will therefore investigate two different cases when varying the resolution: a fixed $\alpha_{\rm crit}$ and a scaled $\alpha_{\rm crit} \propto M_{\rm particle}^{-2/3}$. In the former case higher gas densities (or lower velocity dispersions) are required when the numerical resolution increases, whereas in the latter case gas with the same physical properties is selected regardless of the numerical resolution.

It is instructive to consider two limiting cases. The first is when the turbulent velocity dispersion is negligible compared to the thermal dispersion (i.e. $\sigma_{\rm th} \gg \sigma_{\rm turb}$). In this case, the instability criterion reduces to the thermal Jeans criterion,
\begin{align}
    T &< \frac{\mu m_{\rm H}^{4/3} G}{3 k_{\rm B} X^{1/3}} \alpha_{\rm crit} \langle N_{\rm ngb} \rangle^{2/3} M_{\rm particle}^{2/3} n_{\rm H}^{1/3}, \\
    &< 2.5 \times 10^3~{\rm K} \left( \frac{n_{\rm H}}{1 ~\rm cm^{-3}} \right) \left( \frac{\langle N_{\rm ngb} \rangle}{65} \right)^{2/3} \left( \frac{M_{\rm particle}}{10^5~\rm M_\odot} \right)^{2/3} \alpha_{\rm crit}, \label{eq:jeans}
\end{align}
where we assumed that the mean molecular mass is $\mu=1.3$. This version of the instability criterion gives the maximal temperature that star-forming gas can have as a function of the gas density and the numerical resolution. Equation (\ref{eq:jeans}) shows that the temperature of star-forming gas depends explicitly on the resolution of the simulation. In Fig.~\ref{fig:sfcriteria}, we compare the instability criterion with the equilibrium temperature curves for fixed metallicities. Because of the degeneracy between $\alpha_{\rm crit}$ and $M_{\rm particle}$, the instability criterion is only shown for fixed $M_{\rm particle} \alpha_{\rm crit}^{2/3}$. The figure shows that for a  lower (higher) resolution or larger (smaller) $\alpha_{\rm crit}$ the temperature of the selected gas can be higher (lower) for a fixed gas density.

The second limiting case is when the thermal velocity dispersion is negligible compared to the turbulent dispersion. In this case the instability criterion reduces to a density criterion for a fixed velocity dispersion,
\begin{align}
    n_{\rm H} &> \frac{X \sigma_{\rm turb}^6}{m_{\rm H} \alpha_{\rm crit}^3 G^3 \langle N_{\rm ngb} \rangle^2 M_{\rm particle}^2}, \\
    &> 0.14 ~{\rm cm^{-3}} \left( \frac{\sigma_{\rm turb}}{5~\rm km ~\rm s^{-1}} \right)^6 \left( \frac{\langle N_{\rm ngb} \rangle}{65} \right)^{-2} \left( \frac{M_{\rm particle}}{10^5~\rm M_\odot} \right)^{-2} \alpha_{\rm crit}^{-3}. \label{eq:density-cut}
\end{align}
This effective density criterion is very sensitive to the velocity dispersion, which can be seen in Fig.~\ref{fig:sfcriteria} by comparing differently coloured solid curves. This means that gas at a fixed density quickly becomes ineligible for SF as the turbulent velocity dispersion increases. The dependence on the resolution is very strong for simulations with a fixed $\alpha_{\rm crit}$, but for simulations in which $\alpha_{\rm crit}$ is scaled with the resolution, gas with the same densities is selected.

%% file: initial_conditions.tex
\subsection{Initial conditions}
\defcitealias{springel2005}{S05}
\defcitealias{navarro1997}{NFW}

We simulate an isolated galaxy consisting of an exponential disk of stars and gas embedded in a dark matter halo. To reduce unnecessary computational expense, the dark matter halo is modelled using an external gravitational potential instead of with dark matter particles. 

The initial conditions are generated with the \texttt{MakeNewDisk} code 
(\citealp{springel2005}; hereafter \citetalias{springel2005}). The dark matter halo follows a \citet{hernquist1990} profile, but compared to \citetalias{springel2005} we use a slightly different absolute normalization and a scale radius that produces a better match to a dark matter halo that follows a \citet{navarro1997} (hereafter \citetalias{navarro1997}) profile. We define the \citet{hernquist1990} profile as:
\begin{align}
    \rho_{\rm Hern} (r) &= \frac{M_{\rm Hern}}{2 \pi} \frac{r_\star}{r(r+r_\star)^3},
\end{align}
where $M_{\rm Hern}$ is the total mass obtained when integrating the profile to $r=\infty$ and $r_\star$ is the scale radius (and also the half-mass radius). To set the total halo mass, we use the parameter $M_{200}$, where $M_{200}$ is the mass within $R_{200}$, the radius within which the average density of the halo is 200 times the critical density of the universe. 
Using $M_{200} = \int_0^{R_{200}} \rho_\text{Hern} \, 4\pi r^2\, {\rm d}r$ gives
\begin{align}
    M_{200} &= \frac{M_{\rm Hern} R_{200}^2}{(R_{200} + r_\star)^2}.
\end{align}
This means the \citet{hernquist1990} profile can be written as
\begin{align}
    \rho_{\rm Hern} (r) &= \frac{M_{200} (R_{200} + r_\star)^2}{2\pi R_{200}^2} \frac{r_\star}{r(r+r_\star)^3}.
\end{align}
When we constrain the central profile to follow the \citetalias{navarro1997} profile
\begin{align}
    \rho_{\rm NFW}(r) &= \frac{M_{200}}{\frac{4\pi}{3} R_{200}^3} \frac{1}{3\left( \ln(1+c) - \frac{c}{1+c} \right) \frac{r}{R_{200}} \left( \frac{r_{\rm s}}{R_{200}} + \frac{r}{R_{200}} \right)^2},
\end{align}
we obtain:
\begin{align}
    \left( \frac{r_\star}{R_{200}} \right)^2 &= \frac{2}{c^2} \left( \ln\left(1+c\right) - \frac{c}{1+c} \right) \left( \frac{r_\star}{R_{200}} + 1 \right)^2, \label{eq:findroot}
\end{align}
where $c$ is the \citetalias{navarro1997} concentration parameter. \citetalias{springel2005} made the simplifying assumption that $r_\star/R_{200} \ll 1$ and omitted the factor $(r_\star/R_{200}+1)^2$ from equation~(\ref{eq:findroot}). However, in order to obtain a better match to the \citetalias{navarro1997} profile this assumption is not required. The only positive root of equation (\ref{eq:findroot}) for the scale length is
\begin{align}
    r_\star = \frac{b + \sqrt{b}}{1-b} R_{200},
\end{align}
where $b$ is defined as
\begin{align}
    b &\equiv \frac{2}{c^2} \left( \ln(1+c) - \frac{c}{1+c} \right),
\end{align}
and $R_{200}=(G M_{200}/100 H_0^2)^{1/3}$, where $G$ is the gravitational constant and $H_0$ is the Hubble constant at $z=0$.
In the limit $\sqrt{b}\ll 1$ (which corresponds to large NFW concentrations), we have $b \ll \sqrt{b}$ so that $r_\star \ll R_{200}$. 
and the result of \citetalias{springel2005} is obtained. The calculation of the density distributions and the velocities of the stellar and gas particles remain the same as in \citetalias{springel2005}, but with a differently normalised \citet{hernquist1990} profile.
We set the angular momentum $J$ of the dark matter halo using the dimensionless spin parameter $\lambda \equiv J |E|^{1/2}/ GM_{200}^{5/2}$, where $E$ is the total energy of the dark matter halo.

We choose a virial mass $M_{200} = 1.37 \times 10^{12}~\rm M_\odot$ and a concentration parameter $c=9$ \citep[based on the mass-concentration relation from][using the \citet{planck2016} cosmology]{correa2015}. The dimensionless spin parameter of the halo is set to $\lambda = 0.033$ \citep{oppenheimer2018}, which yields a radial disk scale length of $r_{\rm d} = 4.3~\rm kpc$ (\citealp{mo1998}; \citetalias{springel2005}). The disk mass is $M_{\rm disk}=5.48 \times 10^{10}~\rm M_\odot$, which corresponds to 4 per cent of the virial mass, of which the stellar disk contains 70 per cent, $M_{\rm disk, \star} = 3.836 \times 10^{10}~\rm M_\odot$, with a fixed stellar disk scale-height of 10 per cent of the radial disk scale length, i.e. $0.43~\rm kpc$. The remaining 30 per cent of the disk is gas, corresponding to $M_{\rm disk, gas}=1.644\times 10^{10}~\rm M_\odot$. Using these parameter values the stellar mass is comparable to those of the more massive disk galaxies in observations of KS relations \citep[e.g.][]{bigiel2008,bigiel2010} and we sample gas surface densities up to $\sim 100~\rm M_\odot ~\rm pc^{-2}$ as observed for massive disk galaxies \citep[e.g.][]{walter2008}. In our fiducial model we assume solar metallicity ($Z=Z_\odot =0.0134$; \citealt{asplund2009}), but we investigate the effect of different metallicities in \S \ref{subsec:fid}.

The gas initially has a temperature of $10^4~\rm K$. We assume vertical hydrostatic equilibrium, where the disk scale height of the gas is set by the pressure and the gravity of the gas and stars and therefore depends on the radius. Note that in the centre the initial gas scale height is much smaller than the stellar disk scale height; they only become comparable in the outer region. 

To prevent an artificial collapse of the gas disk and a spurious initial burst of SF, we give a fraction of the stellar particles an age distribution corresponding to a constant SFR of $10~\rm M_\odot ~\rm yr^{-1}$ over the last 100 Myr and allow them to inject CC SN and SNIa feedback.  

The initial conditions are made publicly available as an example in the \texttt{SWIFT} code repository and can be found in the \texttt{IsolatedGalaxy} examples at \href{https://gitlab.cosma.dur.ac.uk/swift/swiftsim}{www.swiftsim.com}. 

%% file: KS_relation.tex
\begin{figure*}
    \includegraphics[width=\textwidth]{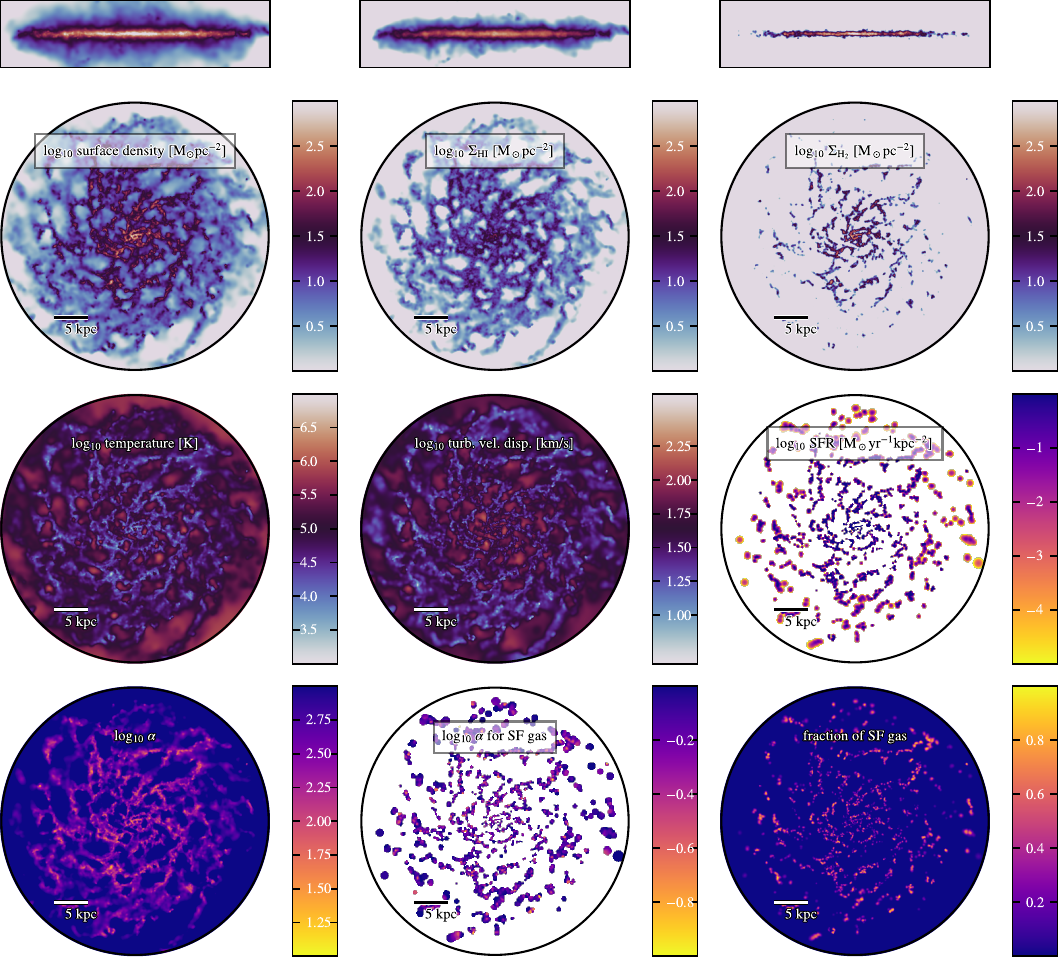} \vspace{-5mm}
    \caption{The fiducial simulation with $Z=Z_\odot$ at $t=1~\rm Gyr$. \emph{Top two rows:}  Edge-on and face-on gas surface density for the total gas (left column), atomic hydrogen (middle column) and molecular hydrogen (right column). \emph{Third row:} Face-on view of the mass-weighted temperature (left), mass-weighted 3D turbulent velocity dispersion (middle), and SFR surface density (right). \emph{Bottom row:} Face-on view of the mass-weighted virial parameter $\alpha$ for all gas (left) and only star-forming gas (middle), and the mass fraction of gas that is star-forming (right). The thickness of the projection is $50~\rm kpc$.}
    \label{fig:Hsurface}
\end{figure*}

\subsection{The fiducial model} \label{subsec:fid}
\defcitealias{bigiel2008}{B08}
\defcitealias{bigiel2010}{B10}
\defcitealias{pessa2021}{P21}
\defcitealias{querejeta2021}{Q21}
\defcitealias{ellison2020}{E20}
\defcitealias{kennicutt1998}{K98}

In this section, we investigate the SF properties of three different simulations that are run using the fiducial model, which uses the gravitational instability criterion for SF (equation~\ref{eq:instability-crit}) with $\alpha_\text{crit}=1$, the \citet{schmidt1959} law for SF (equation~\ref{eq:schmidt-law}) with $\varepsilon=0.01$, and a particle mass of $10^5\,\text{M}_\odot$. We will compare simulations with metallicities of $0.5 Z_\odot$, $Z_\odot$, and $2 Z_\odot$. Before doing so, we first provide a visual impression of the main properties of the galaxy with solar metallicity. 

Fig.~\ref{fig:Hsurface} shows face-on images of different properties of the galaxy in the fiducial simulation with metallicity $Z=Z_\odot$ at time $t=1$~Gyr. The top two rows show the surface densities for all gas (left column), atomic hydrogen (middle column), and molecular hydrogen (right column), where the species fractions are calculated using \texttt{Chimes}. The galaxy spiral structures are most prominent in the molecular hydrogen maps, but the atomic hydrogen also shows clear spiral structure. The SFR surface density (third row, right column) mostly traces the molecular hydrogen surface density. 

Across most of the galaxy, the thermal and turbulent velocity dispersions are similar\footnote{The colour scale of the temperature is identical to the 3D velocity dispersion, so a temperature correspond to the same thermal velocity dispersion as indicated on the colour scale of the 3D velocity dispersion.} (compare the left and middle columns of the third row). The dispersion tends to be small when the surface density is high, and vice versa. However, in the regions with the highest surface densities the mass-weighted virial parameter (bottom left) is still high ($\alpha \gg 1$). Indeed, most of the gas in the disk has a high value of $\alpha$ and is not forming stars (bottom right), while most of the star-forming gas has $\alpha \approx 1$ (bottom centre). 

The instability criterion can be converted into a function of the gas surface density $\Sigma$ and the scale height $h$. The mass of the cloud is given by $M = \frac{4\pi}{3}\rho h^3$ and the density is given by $\rho=\Sigma/h$. This results in
\begin{align}
    \alpha &= \frac{\sigma^2}{G M^{2/3} \rho^{1/3}} = \left( \frac{4\pi}{3} \right)^{-2/3} \frac{\sigma^2}{G \Sigma h}, \\
    &= 7.1 \left( \frac{\Sigma/\sigma^2}{0.4 ~\rm M_\odot ~\rm pc^{-2} ~(\rm km~\rm s^{-1})^{-2}} \right)^{-1} \left( \frac{h}{40~\rm pc} \right)^{-1}.
\end{align}
This means that the ratio $\Sigma/\sigma^2$ is a measure of the boundedness of the gas. \citet{leroy2017} measured $\Sigma/\sigma^2$ for molecular gas on spatial scales from $300~\rm pc$ to $1~\rm kpc$ in M51 and their measurements indicate that the stability parameter ranges between $\alpha\approx 5$ and $\approx 14$, implying that most of the molecular gas in the ISM is gravitationally stable. Not surprisingly, these measurements are lower than our mass weighted values that include all gas. The molecular gas mass weighted values  are in good agreement with these observations and range between $\alpha \approx 2$ and $20$ (not shown).

\subsubsection{Star fromation histories}
\begin{figure}
    \includegraphics[width=\columnwidth]{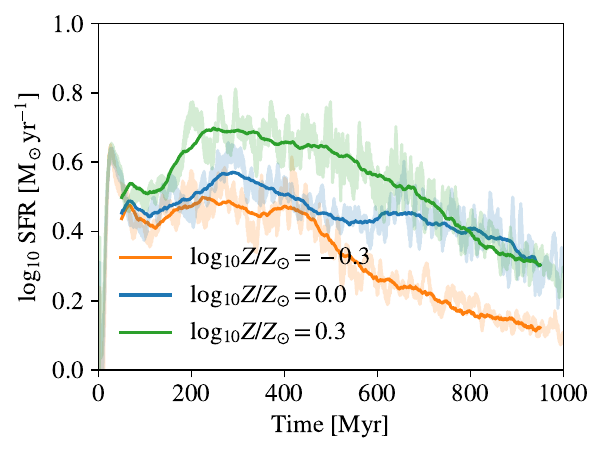} \vspace{-7mm}
    \caption{Comparison of the star formation histories for the simulations with different metallicities (different colours). The thick lines show the 100 Myr moving average of the SFR and the transparent lines show the SFR at the time resolution of the simulation. The differences are small but the SFR is higher (lower) for simulations with higher (lower) metallicities. The SFR varies by $\approx 0.2$ dex over short time scales.}
    \label{fig:SFHmetal}
\end{figure}

Fig.~\ref{fig:SFHmetal} shows the star formation history (SFH) for the simulations with different metallicities. The thick lines show the SFH averaged over\footnote{We average the instantaneous SFR that is logged every single time step over 100 Myr, i.e. SFR$_{\rm avg}=\sum_{t'=t-50~\rm Myr}^{t'=t+50~\rm Myr} {\rm SFR}(t')\Delta t(t') / 100~\rm Myr$.} 100 Myr, we average the instantaneous SFR that is logged every single time step over 100 Myr. The different simulations are within 0.4 dex of each other and lower metallicity results in a lower SFR, in agreement with the findings of \citet[][]{richings2016} for a different SF criterion. Besides the 100 Myr averaged SFH we also show the instantaneous SFR that is updated every time a gas particle is active and therefore has the same time resolution as the simulation. Here the SFH varies by about 0.2 dex over times scales $\sim 10 ~\rm Myr$. The small artificial initial SF peak at $t<50~\rm Myr$ does not impact the results at $t> 100 ~\rm Myr$ because we include feedback from existing stars. After a few 100 Myr, the SFR declines because the gas in the disk is being ejected and consumed.

\begin{figure*}
    \includegraphics[width=\textwidth]{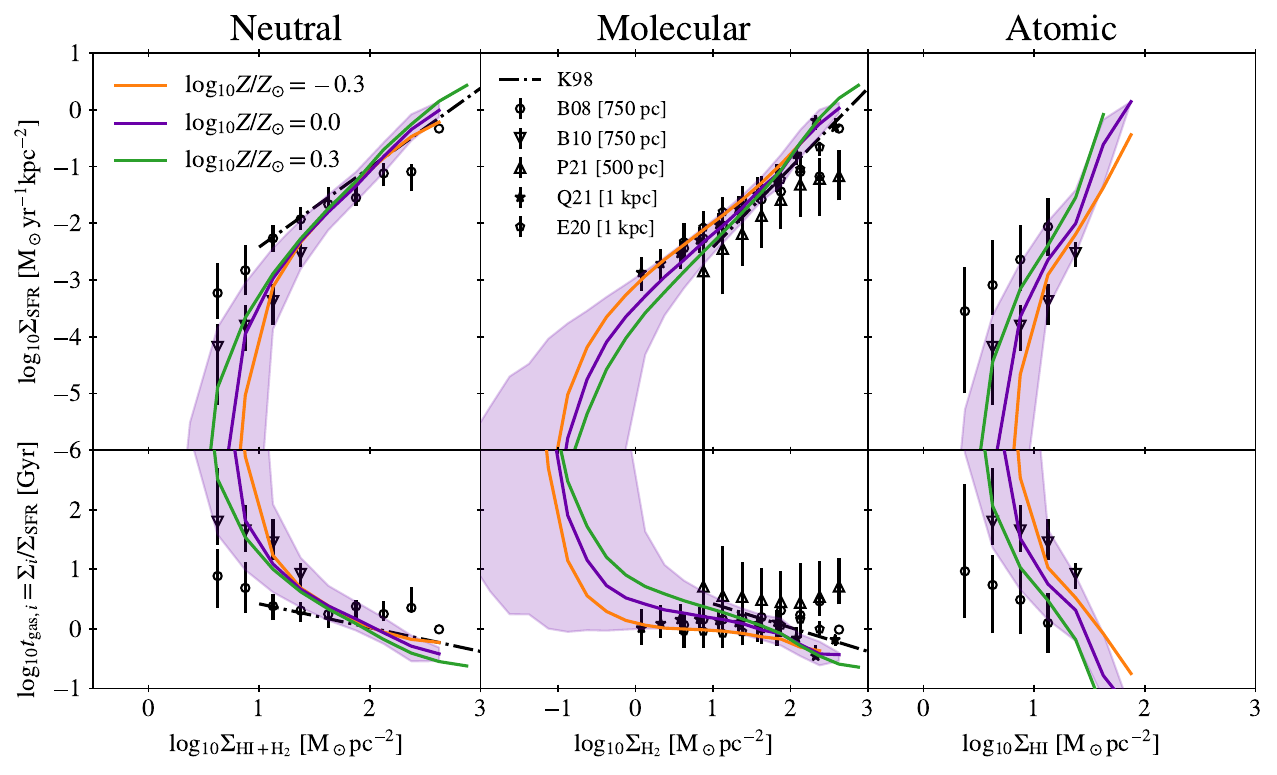} \vspace{-7mm}
    \caption{Comparison of the (face-on) spatially-resolved median star formation rate surface density (i.e., the Kennicutt-Schmidt (KS) relation) (top) and the median gas consumption time $t_{\rm gas}$ (bottom) as a function of gas surface density for neutral (left), molecular (middle) and atomic (right) hydrogen. The results are averaged over cells of size $750~ \rm pc$ for different metallicities (different colours). The shaded region shows the 16\textsuperscript{th} and 84\textsuperscript{th} percentile scatter for the $Z=Z_\odot$ simulation. We compare with observations of galaxy-averaged KS relations \citepalias{kennicutt1998}, spatially-resolved atomic and molecular hydrogen surface densities \citepalias{bigiel2008,bigiel2010}, and spatially-resolved molecular hydrogen surface densities \citepalias{pessa2021,querejeta2021,ellison2020}. The error bars show the 16\textsuperscript{th} and 84\textsuperscript{th} percentile scatter. The simulations are in good agreement with the data. A higher metallicity shifts the cut-off in the neutral and atomic KS relations to lower lower surface densities, while the cut-off in the molecular KS relation shifts to higher surface densities.}
    \label{fig:KS-neutral}
\end{figure*}

\subsubsection{The spatially-resolved Kennicutt-Schmidt relation}
The SFH in general does not provide a stringent test of models of SF \citep[e.g.][]{schaye2010,hopkins2013} because it is determined mostly by the gas supply and feedback processes on galaxy scales. Galaxy-averaged KS relations have the same problem because they are effectively SFRs that are re-normalized to the galaxy size. More useful tests are the spatially-resolved Kennicutt-Schmidt (KS) relations \citep[e.g.][]{kennicutt2007}, between gas surface density and SF,
\begin{align}
    \Sigma_{\rm SFR} &= A \Sigma_i^N,
\end{align}
where $\Sigma_{\rm SFR}$ is the SFR surface density in a spatially-resolved region, $A$ is the normalisation constant, $\Sigma_i$ is the surface density of gas in the same region where $i$ is the component (\ion{H}{I}, $\rm H_2$ or \ion{H}{I} + $\rm H_2$) and $N$ is the slope of the power-law relation.

Focusing again on Fig.~\ref{fig:Hsurface}, we show the total, atomic and molecular gas surface densities for the \texttt{fid} simulation in the top panel. We see that most of the disk area consists of atomic hydrogen. Molecular hydrogen is largely confined to the highest-density regions. Comparing the top right and centre right panels of Fig.~\ref{fig:Hsurface}, we see that the molecular hydrogen surface density is spatially closely correlated with the SFR surface density.

The top three panels of Fig.~\ref{fig:KS-neutral} show the spatially-resolved KS relations for, respectively, neutral, molecular and atomic gas. The different coloured curves show the median relations in 0.25-dex-wide surface density bins for our three different metallicities. The shaded regions indicate the 16\textsuperscript{th} and 84\textsuperscript{th} percentiles for the case of solar metallicity. The relations are averaged over square regions of size $750~\rm pc$, matching the typical resolution of observations. In \S~\ref{sec:azi_avg} we will also consider the azimuthally averaged KS relation (i.e.\ the KS relation averaged over radial rings of width $\Delta r \approx 750 ~\rm pc$). 

Data points with error bars indicate the median and the 16\textsuperscript{th}-84\textsuperscript{th} percentile scatter for observations of galaxies that have metallicity variations similar to our simulations. None of the relations includes a correction for helium and heavy elements in the surface densities. We compare to observations of global, galaxy disk-averaged KS relations in spiral and star-forming galaxies (\citealt{kennicutt1998}; hereafter \citetalias{kennicutt1998}), of spatially resolved atomic and molecular hydrogen surface densities (\citealt{bigiel2008}; hereafter \citetalias{bigiel2008}; \citealt{bigiel2010}; hereafter \citetalias{bigiel2010}), and of spatially resolved molecular hydrogen surface densities (\citealt{ellison2020}; hereafter \citetalias{ellison2020};\citealt{pessa2021}; hereafter \citetalias{pessa2021}; \citealt{querejeta2021}; hereafter \citetalias{querejeta2021}). We corrected the surface densities of \citetalias{bigiel2008}, \citetalias{bigiel2010}, \citetalias{ellison2020} and \citetalias{querejeta2021} by removing the factor for helium and heavy elements. 

We note that there is considerable uncertainty in the observations due to the use of CO as a tracer for molecular hydrogen. First, the $\alpha_{\rm CO}$ factor is assumed to be constant but likely depends on metallicity (e.g.\ $\propto Z^{-1}$; \citealt{narayanan2012}). If gas with higher $\Sigma_i$ has higher metallicity, then this would imply that the KS relation is steeper (and hence that the $t_{\rm gas}$ relation has a more negative slope). Second, gas with densities below the critical density of CO ($n_{\rm crit, CO}=10^3~\rm cm^{-3}$; \citealt{schoier2005}) is only detected when the clouds have large optical depths $\tau$ such that the effective critical density is $n_{\rm crit eff, CO} = n_{\rm crit, CO}/\tau$. This means that CO observations can underestimate the CO mass. If the volumetric density increases with the surface density, then correcting for this effect would again steepen the KS relation. Third, the $\alpha_{\rm CO}$ factor may vary due to differences in the optical depth of the gas caused by higher turbulent velocities \citep[e.g.][]{teng2022}. Fourth, the $\alpha_{\rm CO}$ factor is smaller inside the centre of galaxies and therefore the slope of the KS relation will steepen \citep{denBrok2023}. Fifth, the commonly adopted value for the ratio of CO (2-1)/(1-0) is 0.65 but observed ratios range from 0.3 to 1.0 \citep{denbrok2021}. There are indications of a trend between this ratio and $\Sigma_{\rm SFR}$ \citep{denbrok2021,yajima2021}. Similarly, the CO (3-2) transition ratios are not constant \citep{leroy2022}. This might be because of different densities and temperatures in the CO gas, which implies uncertainty in the $\alpha_{\rm CO}$ factor.
  
In Fig.~\ref{fig:KS-neutral} the agreement between the data and the simulations is good, both for the median and the scatter, and both in terms of the normalizaton and the slope. The neutral and atomic KS relations agree better with \citetalias{bigiel2010} (\citetalias{bigiel2008}) at low (high) surface densities as would be expected because \citetalias{bigiel2010} (\citetalias{bigiel2008}) observed the outer (inner) parts of the same galaxies. Additionally, below $\Sigma_{\ion{H}{I} + \rm H_2}\approx 10~\rm M_\odot ~\rm pc^{-2}$, \citetalias{bigiel2008} detected much fewer regions than \citetalias{bigiel2010}. The neutral and atomic KS relations decline sharply below the canonical SF threshold of $\Sigma_i=10 ~\rm M_\odot ~\rm pc^{-2}$ \citep[e.g.][and references therein]{schaye2004}. At high $\Sigma_{\rm \ion{H}{I}+H_2}$ the slope approaches the canonical value $N\approx 1.4$  \citep{kennicutt1998}. The slope of the atomic KS relation is very steep, $N\approx 3$, consistent with the observations. For the remainder of this paper, we will use the gas consumption time $t_{\rm gas}$ (bottom of Fig. \ref{fig:KS-neutral}) rather than the KS relation, because a time scale can be interpreted physically and because it reduces the dynamic range of the y-axis, which better highlights differences. We call this the $t_{\rm gas}$ relation. 

Increasing the metallicity from $\log_{10} Z/ Z_\odot = -0.3$ (orange) to $0.3$ (green) barely impacts the gas consumption time for neutral gas, decreases (increases) that for atomic (molecular) gas, particularly at low surface densities. This mainly reflects the higher molecular fractions for higher metallicities due to the increased abundance of dust grains that catalyse molecule formation and shield dissociating radiation, and the increase in metal cooling that enables more gas to cool to the low temperatures required for molecule formation.  
Despite these differences, all metallicity variations fall within the range spanned by the observations (which, as discussed above, may however suffer from a metallicity-dependent bias). We now turn to the analysis of the effect of the spatial averaging method.

\begin{figure*}
    \includegraphics[width=\textwidth]{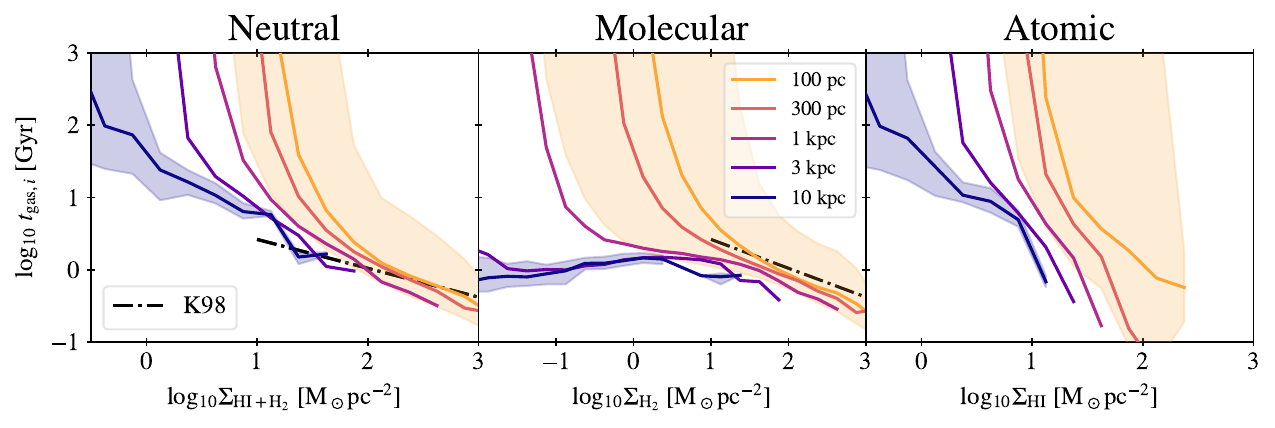} \vspace{-7mm}
    \caption{Comparison of the spatially-resolved relation between the median gas consumption time and the gas surface density for neutral (left), molecular (centre) and atomic hydrogen (right), averaged over different spatial scales (different coloured lines) for the fiducial simulation with $Z=Z_\odot$. For the largest (10 kpc) and lowest (100 pc) bin sizes the shaded regions indicate the 16\textsuperscript{th} and 84\textsuperscript{th} percentile scatter. At low surface densities the relations steepen. For smaller (larger) observed patches, this break in the $t_{\rm gas}$ relation becomes more (less) pronounced and shifts to higher (lower) surface densities. For the largest spatial patches, there is no break.}
    \label{fig:spatialres}
\end{figure*}

\subsubsection{The spatial averaging scale}

Fig.~\ref{fig:spatialres} shows the neutral, molecular and atomic $t_{\rm gas}$ relations for the fiducial simulation averaged over different spatial scales. For spatial bins of size $\leq 3~\rm kpc$, the neutral and atomic gas consumption times steepen rapidly at low surface densities, indicating very inefficient SF. However, for larger spatial bin sizes this break moves to smaller surface densities, which agrees with observations \citep[e.g.][]{onodera2010}. For the largest bin size there is no break at all and the gas consumption time is reasonably well approximated by a power law of the surface density. This indicates that the larger spatial bins contain subregions with higher surface density and smaller, local gas consumption times. For a given spatial bin size, the break occurs at smaller molecular than atomic surface densities. As expected, the scatter in $t_{{\rm gas},i}$ decreases when averaging over larger spatial scales. At the largest spatially-averaging scales, the $t_{{\rm gas}, \ion{H}{I} + \rm H_{2}}$ relation is close to the observations of \citetalias{kennicutt1998} for entire galaxies (dot-dashed lines). The normalisation of the $t_{\rm gas}$ relations are lower for larger spatially-averaging scales, which is in agreement with the different normalisations of the galaxy-averaged KS relation of \citet{kennicutt1998} and the spatially-resolved KS relation of \citet{kennicutt2007} who found a difference in normalisation similar to what we find. 

\defcitealias{Schruba2011}{S11}

\begin{figure*}
    \includegraphics[width=\textwidth]{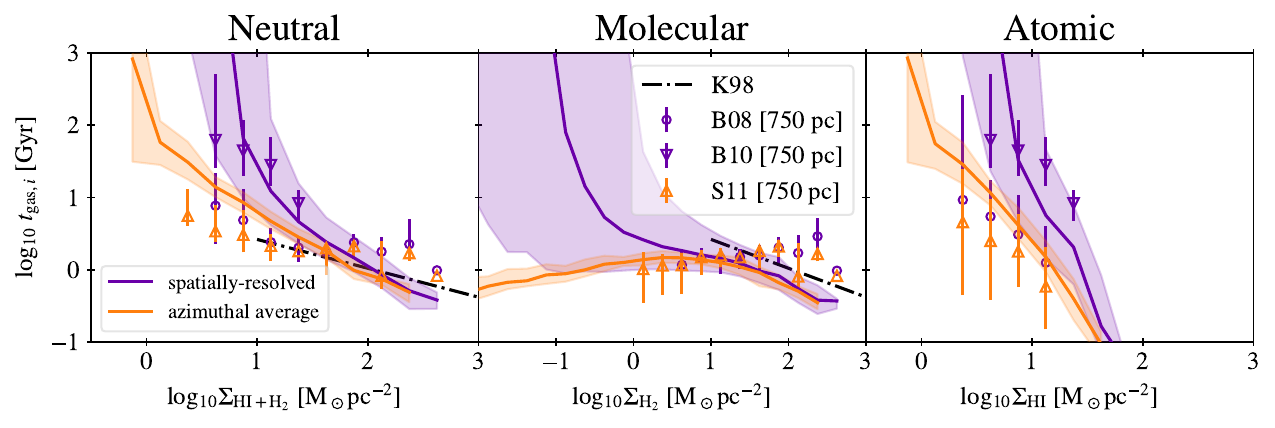} \vspace{-7mm}
    \caption{Comparison of the spatially-resolved (purple) and azimuthally-averaged (orange)  median $t_{\rm gas}$ relations for all neutral (left), molecular (centre) and atomic hydrogen (right) for the fiducial simulation with solar metallicity. The shaded regions indicate the 16\textsuperscript{th} - 84\textsuperscript{th} percentile scatter. We compare with disk-averaged observations of spiral and star-forming galaxies (dot-dashed line; \citetalias{kennicutt1998}) azimuthally-averaged observations of \citetalias{Schruba2011} and the spatially-resolved data from \citetalias{bigiel2008} and \citetalias{bigiel2010}. The error bars show the 16\textsuperscript{th}-84\textsuperscript{th} percentile scatter. The observations and simulations use the same spatial and azimuthal resolution of $750~\rm pc$. The spatially-resolved relations are much steeper than the azimuthally-averaged ones, particularly for neutral and atomic gas. The simulations agree with both types of observed relations.}
    \label{fig:radialvsspatial}
\end{figure*}

\subsubsection{Azimuthally-averaged vs.\ spatially-resolved $t_\text{gas}$ relations} \label{sec:azi_avg}
Fig.~\ref{fig:radialvsspatial} shows a comparison between the spatially-resolved (i.e. local) and the azimuthally-averaged $t_{\rm gas}$ relations for bin sizes of $750~\rm pc$ and for neutral, molecular and atomic hydrogen, respectively. The predicted spatially-resolved relations are compared with spatially-resolved observations \citepalias{bigiel2008,bigiel2010} and the predicted azimuthally-averaged relations are compared with azimuthally-averaged observations (\citealt{Schruba2011}; hereafter \citetalias{Schruba2011}). We corrected the surface densities of \citetalias{Schruba2011} by removing the factor for helium and heavy elements. For reference, we also show the observed galaxy-averaged $t_{{\rm gas},i}$ \citepalias{kennicutt1998}. The simulations are overall in good agreement with the observations for the different phases and averaging methods. Compared to \citetalias{Schruba2011} the predicted gas consumption times are slightly too large at low surface densities, but \citetalias{Schruba2011} only use data from the inner regions of galaxies (i.e.\ mostly the same data as \citetalias{bigiel2008}, and partly \citetalias{bigiel2010}). 

While the spatially-resolved and azimuthally-averaged gas consumption time scales converge at high surface densities, they differ strongly at low surface densities because the spatially-resolved $t_{\rm gas}$ relations are steeper. Comparing Figs~\ref{fig:spatialres} and \ref{fig:radialvsspatial}, we see that the azimuthally-averaged relations using a radial bin size of 750~pc are similar to the spatially-resolved relations using a bin size of $\sim 3-10$~kpc, which is comparable to the radial disk scale length of 4.3~kpc.

For the remainder of the paper, we will focus on the neutral and molecular $t_{\rm gas}$ relations. We now turn to resolution tests.

\begin{table}
\begin{center}
\caption{Resolution variations
\label{tbl:res}}
\begin{tabular}{lrrl}

\hline
  Simulation & particle mass & softening & $\alpha_{\rm crit}$ \\
             & ($\rm M_\odot$) & ($\rm pc$) &    \\
\hline
  \texttt{fid}       & $10^5$ & 200 & 1.0  \\
  \texttt{highres8fixedalpha}       & $1.25\times 10^4$ & 100 & 1.0  \\
  \texttt{lowres8fixedalpha}       & $8\times 10^5$ & 400 & 1.0  \\
  \texttt{lowres64fixedalpha}       & $6.4\times 10^6$ & 800 & 1.0  \\
  \texttt{lowres512fixedalpha}       & $5.12\times 10^7$ & 1600 & 1.0  \\
  \texttt{highres8scaledalpha}       & $1.25\times 10^4$ & 100 & 4.0  \\
  \texttt{lowres8scaledalpha}       & $8\times 10^5$ & 400 & 0.25  \\
  \texttt{lowres64scaledalpha}       & $6.4\times 10^6$ & 800 & 0.00625  \\
  \texttt{lowres512scaledalpha}       & $5.12\times 10^7$ & 1600 & 0.001562  \\
\hline
\end{tabular}
\end{center}
\end{table}

\subsubsection{Resolution tests}

\begin{figure}
    \includegraphics[width=\columnwidth]{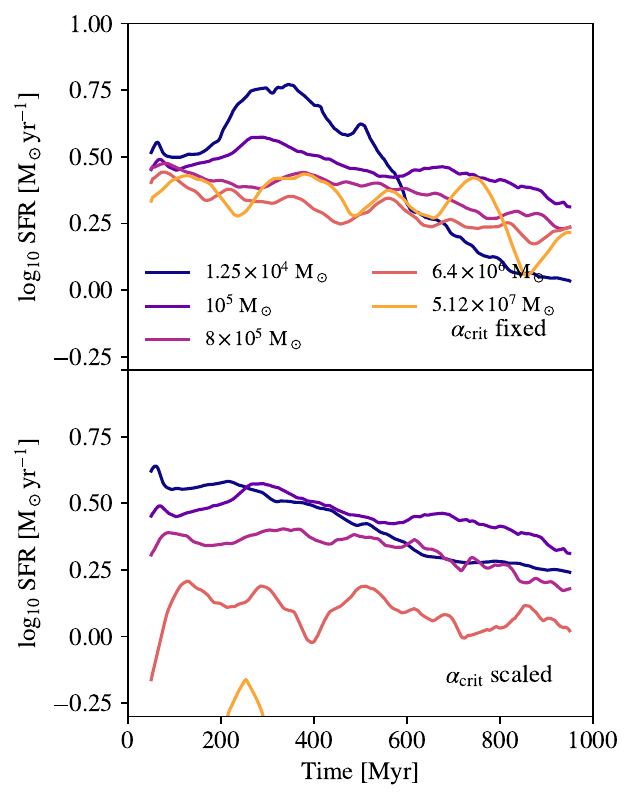} \vspace{-7mm}
    \caption{Comparison of the SFHs for simulations with different resolutions (different colours) and either constant $\alpha_{\rm crit}$ (top panel) or scaled $\alpha_{\rm crit}$ ($\alpha_\text{crit}\propto M_\text{particle}^{-2/3}$; bottom panel). The $M_\text{particle}=10^5~\rm M_\odot$ simulation is the same in both panels. The lines show the 100 Myr moving average SFR. For scaled $\alpha_{\rm crit}$ the SFR is systematically lower for lager particle masses, while for a fixed $\alpha_{\rm crit}$ the results are insensitive to the resolution.}
    \label{fig:SFHres}
\end{figure}

Equation (\ref{eq:instability-crit}) indicates that the selection of star-forming gas depends directly on the numerical resolution. This means that for a fixed value of $\alpha_{\rm crit}$ different regions in ($T, n_{\rm H}, \sigma_{\rm turb}$) space are selected for different resolutions. For a fixed $\alpha_{\rm crit}$ but higher (lower) resolution, gas is selected with higher (lower) densities, lower (higher) temperatures and lower (higher) turbulent velocity dispersions. We therefore look at the convergence with resolution in two different settings: a fixed $\alpha_{\rm crit}$ and a scaled $\alpha_{\rm crit}$ ($\alpha_{\rm crit} \propto M_{\rm particle}^{-2/3}$). For fixed $\alpha_{\rm crit}$ the physical conditions need to be such that the minimum resolved mass is unstable, while for scaled $\alpha_{\rm crit}$ the same mass is required to be unstable when the resolution changes and hence the same regions in ($T, n_{\rm H}, \sigma_{\rm turb}$) space are selected. In Table~\ref{tbl:res} we list the different simulations used for the resolution tests.

Fig.~\ref{fig:SFHres} shows the SFHs for the different resolution variations. We do not expect exact convergence because the initial density profile is stochastically realized and both SF and stellar feedback are implemented stochastically. The SFHs converge for the simulations with fixed $\alpha_{\rm crit}$ as compared to those with a scaled $\alpha_{\rm crit}$. For the higher-resolution simulations the differences in the SFHs are within 0.2 dex, similar to the impact of varying the metallicity by a factor of two (see Fig. \ref{fig:SFHmetal}) and they do not vary systematically with the resolution. The convergence is good for all particle masses for fixed $\alpha_{\rm crit}$ and not good for scaled $\alpha_{\rm crit}$. For scaled $\alpha_{\rm crit}$, simulations with even lower resolutions predict lower SFRs (the simulation with $5.12\times 10^7~\rm M_\odot$ and scaled $\alpha_{\rm crit}$ does only form a few stars). At low resolutions the scaled $\alpha_{\rm crit}$ simulations have the problem that the density of gas that becomes star-forming (following equation~\ref{eq:density-cut}) is much higher than $m_{\rm gas}/ \epsilon^3$, the density at which gravitational softening becomes important.

\begin{figure}
    \centering
    \includegraphics[width=\columnwidth]{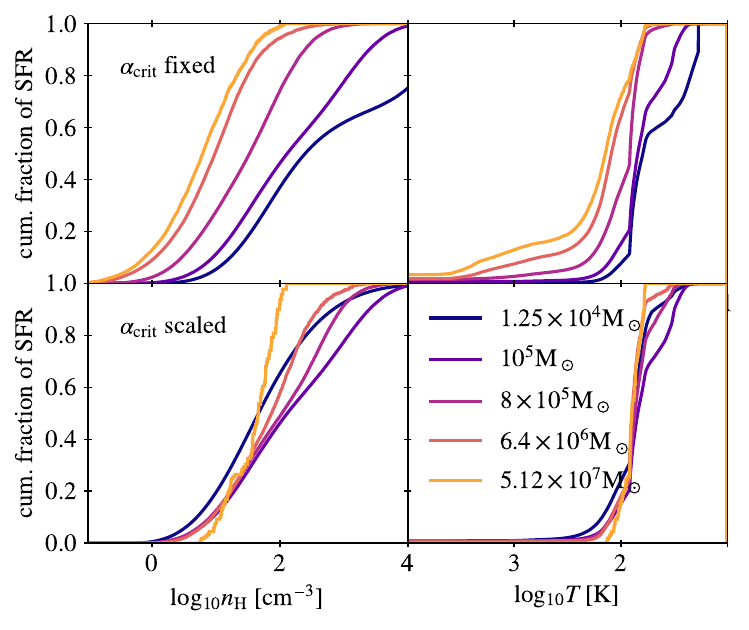}\vspace{-5mm}
    \caption{Comparison of the cumulative fraction of the SFR as a function of the gas density (left) and temperature (right) for simulations with different resolutions (different colours) and either constant $\alpha_{\rm crit}$ (top) or scaled $\alpha_{\rm crit}$ ($\alpha_\text{crit}\propto M_\text{particle}^{-2/3}$; bottom). Because of the explicit resolution dependence of the instability criterion for fixed $\alpha_{\rm crit}$, the convergence is much better if $\alpha_{\rm crit}$ is scaled with the resolution.}
    \label{fig:density-sf-gas}
\end{figure}

Fig.~\ref{fig:density-sf-gas} shows the cumulative fraction of the SFR\footnote{Calculated as $\sum_i {~\rm SFR}(n_{{\rm H},i}<n_{\rm H})/ {~\rm SFR}$ and $\sum_i {~\rm SFR}(T_i>T)/ {~\rm SFR}$.} as a function of density and temperature. As expected, in simulations using a fixed $\alpha_{\rm crit}$ stars form in gas with higher densities and lower temperatures if the resolution is increased. The simulations with scaled $\alpha_{\rm crit}$ converge better in terms of the densities and temperatures of the gas from which the stars are predicted to form because they select gas with the same physical properties. However, Fig.~\ref{fig:SFHres} shows that better convergence of the densities and temperatures of the gas from which stars form does not imply better convergence of the SFR. 

\begin{figure}
    \includegraphics[width=\columnwidth]{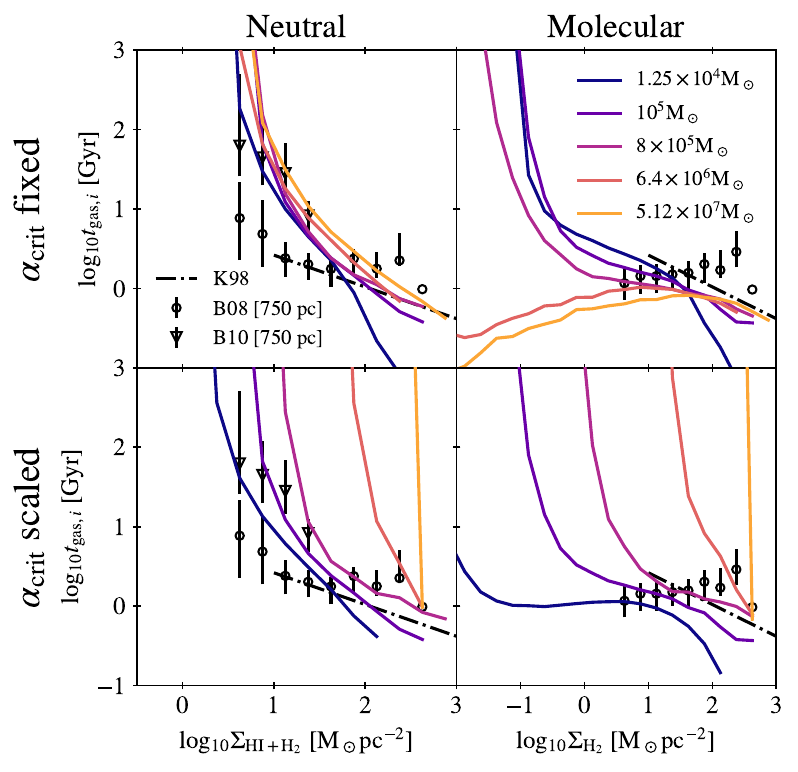} \vspace{-5mm}
    \caption{Comparison of the spatially resolved (750 pc) neutral (left) and molecular (right) $t_{\rm gas}$ relations for simulations with different resolutions (different colours) and different instability criteria (top and bottom). For reference, the dot-dashed line shows the galaxy-averaged $t_\text{gas}$ relation from \citetalias{kennicutt1998} and the data points show the spatially-resolved $t_\text{gas}$ relations from \citetalias{bigiel2008} and \citetalias{bigiel2010}. The neutral $t_{\rm gas}$ relation is close to converged with the numerical resolution for fixed $\alpha_{\rm crit}$, but if $\alpha_{\rm crit}$ is scaled with the resolution, then the relation shifts to longer gas consumption time scales if the resolution is decreased. For the molecular $t_{\rm gas}$ relation the convergence is poor at low surface densities, with higher resolution resulting in longer gas consumption time scales, but at high surface densities ($\Sigma_{\text{H}_2} > 10 \,\text{M}_\odot\, \text{pc}^{-2}$) the results are less sensitive to the resolution.}
    \label{fig:KS-resolution}
\end{figure}

The stellar birth densities and temperatures in the simulations are not particularly interesting properties because they are not observable. In reality, stars form at much higher densities, which remain completely unresolved in the simulations. In fact, if the physical properties of the gas particles that are converted into stars were converged, then increasing the resolution would not result in more realistic properties of star-forming gas and hence would not open new avenues to test the models such as predictions for observational tracers of gas with higher densities. Our goal is therefore not to obtain converged results for the properties of resolution elements that are being converted into stars, but to obtain converged predictions for observables probing fixed physical scales, such as the KS relations.

Fig.~\ref{fig:KS-resolution} shows how the spatially-resolved gas consumption time scales vary with the surface density for the neutral (left) and molecular (right) gas for different numerical resolutions (different colours) and the two $\alpha_{\rm crit}$ variations (different line styles). For fixed $\alpha_{\rm crit}$ the convergence of the neutral $t_{\rm gas}$ is good (within 0.2 dex). The scaled $\alpha_{\rm crit}$ performs less well: the $t_{\rm gas}$ relation shifts to longer consumption time scales if the resolution decreases. At low surface densities the molecular $t_{\rm gas}$ relation shows poor convergence for both fixed $\alpha_{\rm crit}$ and scaled $\alpha_{\rm crit}$. However, for fixed $\alpha_{\rm crit}$ the convergence is reasonable for $\Sigma_{\text{H}_2} > 10 \,\text{M}_\odot\, \text{pc}^{-2}$.

\subsubsection{The $t_\text{gas}$ relations for different gas phases}

\begin{figure}
    \includegraphics[width=\columnwidth]{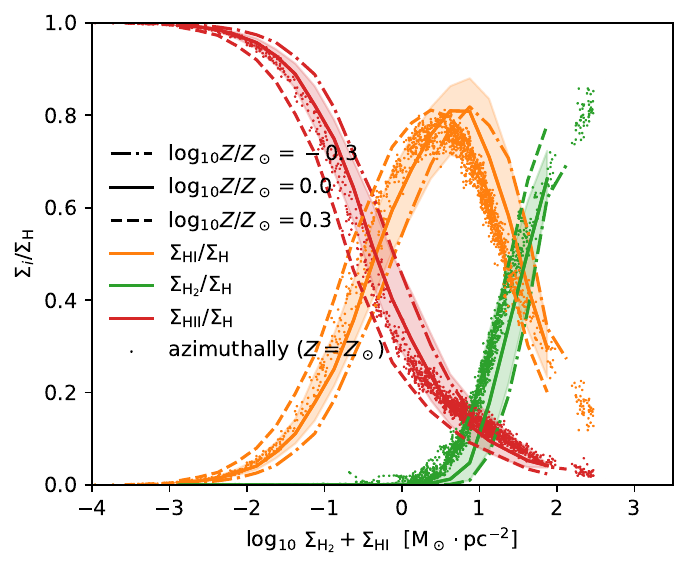} \vspace{-5mm}
    \caption{The contributions of different species of hydrogen (different colours) to the total (face-on) hydrogen surface densities as a function of the neutral hydrogen surface density for different metallicities (different line styles). The lines show the medians and the shaded regions show the 16\textsuperscript{th} and 84\textsuperscript{th} percentile scatter for the case of solar metallicity, averaged over 750~pc regions.  The individual points show the azimuthally averaged values, using 750~pc wide rings. At the highest surface densities, $\Sigma_{\text{H}_2+\ion{H}{I}}\sim 10^2\,\text{M}_\odot\, \text{pc}^{-2}$, the contribution from ionized hydrogen is still not fully negligible. At low surface densities, $\Sigma_{\text{H}_2+\ion{H}{I}} \lesssim 1\,\text{M}_\odot\, \text{pc}^{-2}$, it is possible to ignore the contribution from molecular hydrogen for spatially resolved and azimuthally averaged measurements. The transitions shift to higher surface densities for lower metallicities.}
    \label{fig:diff-species}
\end{figure}

Fig.~\ref{fig:diff-species} shows the hydrogen surface density fractions of ionized, atomic and molecular versus the neutral hydrogen surface density. Lines and data points indicate, respectively, locally and azimuthally averaged quantities, both using bins of $750 ~\rm pc$ at our fiducial resolution. As the neutral surface density increases, the dominant species changes from ionized to atomic to molecular. The transitions are at slightly lower surface densities for the azimuthally averaged case and they shift to lower surface densities if the metallicity is increased. We note that the contribution of ionised hydrogen, which is typically ignored, is significant at all neutral surface densities. Even at our highest densities, $\Sigma_{\ion{H}{I} +\rm H_2}\sim 10^2~\rm M_\odot ~\rm pc^{-2}$, it still accounts for nearly 10 per cent of the gas mass.

\begin{figure}
    \includegraphics[width=\columnwidth]{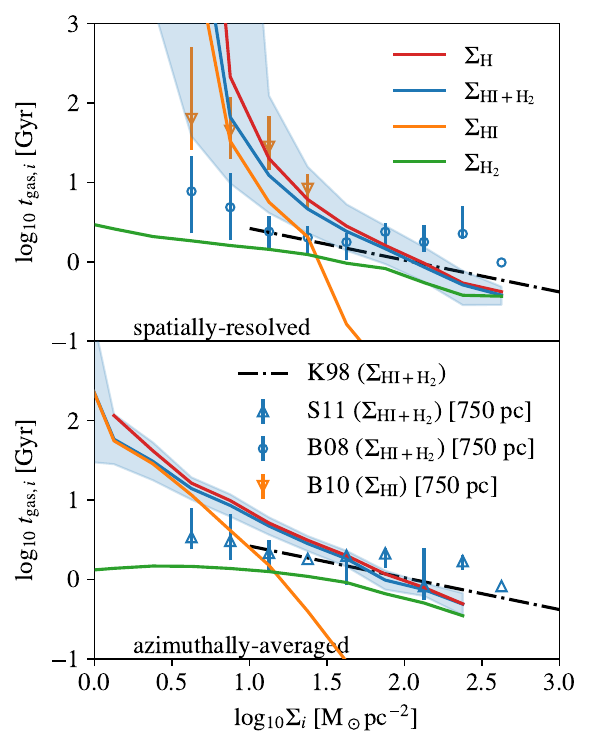}\vspace{-7mm}
    \caption{Comparison of the median $t_{\rm gas}$ relations for different gas phases (different colours) at scales of $750~\rm pc$ for the fiducial model (note that many of the lines were already shown in Fig.~\ref{fig:radialvsspatial}). The shaded regions show the 1 sigma scatter for the neutral gas relation. The top and bottom panels show, respectively, spatially resolved and azimuthally averaged measurements. The simulations are compared with observations of disk averaged relations in spiral and star-forming galaxies (dot-dashed line; \citetalias{kennicutt1998}) and with spatially-resolved observations for neutral gas and atomic gas (top; data points with error bars indicating 16\textsuperscript{th} and 84\textsuperscript{th} percentile scatter; \citetalias{bigiel2008};\citetalias{bigiel2010}) and azimuthally-averaged neutral gas observations (bottom;  error bars indicating 16\textsuperscript{th} and 84\textsuperscript{th} percentile scatter; \citetalias{Schruba2011}). While the molecular gas consumption time scale is nearly constant at $\approx 1$~Gyr, for atomic, neutral and total hydrogen $t_\text{gas}$ is typically larger and increases rapidly towards lower surface densities, particularly for the case of spatially resolved measurements.}
    \label{fig:KS-importance-HII}
\end{figure}

Fig.~\ref{fig:KS-importance-HII} directly compares the $t_{\rm gas}$ relations for the three different hydrogen species and their sum for the fiducial model, this includes some of the lines from Fig. \ref{fig:radialvsspatial}. We show both the spatially resolved (top panel) and the azimuthally averaged $t_{\rm gas}$ relation (bottom panel). The neutral and total hydrogen $t_{\rm gas}$ relations are very close, particularly at high surface densities ($\Sigma_{\ion{H}{I} + \rm H_2} \gg 10~\rm M_\odot ~\rm pc^{-2}$). The atomic $t_{\rm gas}$ relation lies below both and deviates particularly strongly for $\Sigma_{\ion{H}{I}} \ga 10^{1.5}~\rm M_\odot ~\rm pc^{-2}$. The molecular relation deviates the most, giving much shorter $t_{\rm gas}$ for surface densities $\Sigma_{\rm H_2} \la 10^{1.5}~\rm M_\odot ~\rm pc^{-2}$. While for molecular gas the consumption time scale is about 1~Gyr across the full range of surface densities, for the other species the gas consumption time scale increases rapidly with decreasing surface density and this trend is stronger for spatially-resolved than for azimuthally-averaged measurements.

\begin{figure}
    \includegraphics[width=\columnwidth]{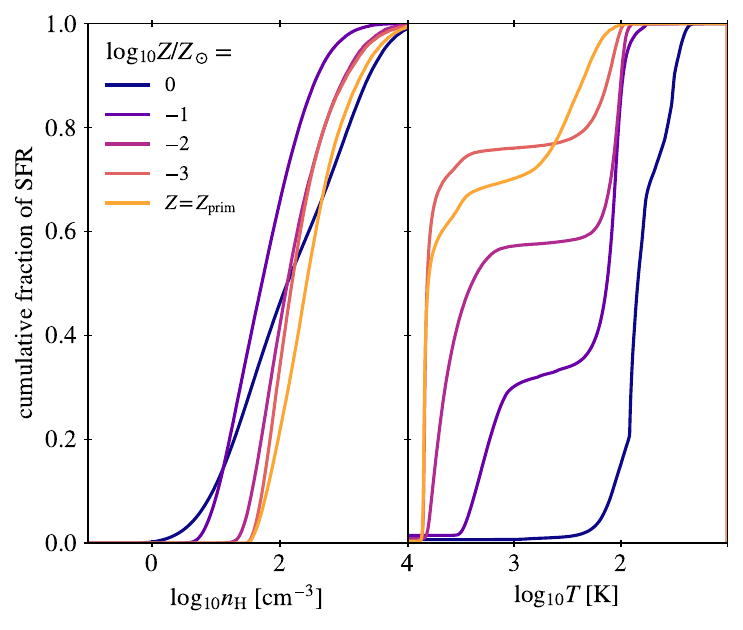} \vspace{-7mm}
    \caption{Comparison of the cumulative contribution of gas with different hydrogen number densities (left) and temperatures (right) to the total SFR for simulations assuming different metallicities (different colours). For lower metallicities star formation occurs in gas with higher temperatures and lower densities. This can be understood by comparing a gravitational instability line with the thermal equilibrium lines for different metallicities in the temperature-density diagram shown in Fig.~\ref{fig:sfcriteria}.}
    \label{fig:Z-temps}
\end{figure}

\begin{figure}
    \includegraphics[width=\columnwidth]{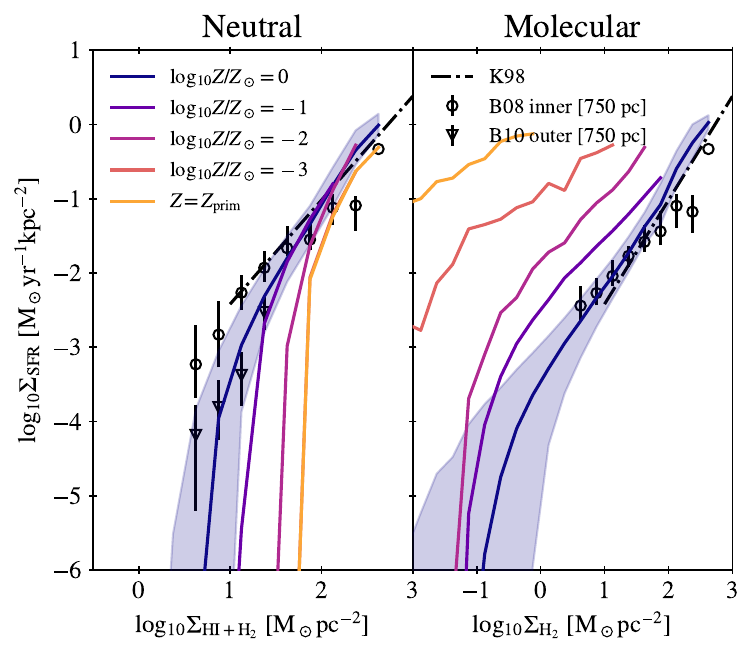} \vspace{-7mm}
    \caption{Comparison of the neutral (left) and molecular (right) spatially-resolved ($750~\rm pc$) median KS relations for simulations assuming different metallicities (different colours). The shaded regions show the 1 sigma scatter for the case of solar metallicity. For reference, the disk-averaged $t_{\rm gas}$ relation from \citetalias{kennicutt1998} (dot-dashed line) and the spatially-resolved $t_{\rm gas}$ relations from \citetalias{bigiel2008} and \citetalias{bigiel2010} are shown. The cut-off in the neutral KS relation shifts to about 1 dex higher surface density going from solar metallicity to primordial abundances, but the relations converge at high surface densities. The molecular KS relation changes dramatically, with lower metallicities giving higher $\Sigma_\text{SFR}$ at a fixed molecular surface density.}
    \label{fig:KS-importance-Z}
\end{figure}

\subsubsection{The impact of gas metallicity}
Fig.~\ref{fig:sfcriteria} showed that at a fixed density, lower metallicity gas has higher equilibrium temperatures. Therefore, for a fixed turbulent velocity dispersion the gravitational instability criterion applied on a fixed mass scale (i.e., for a fixed numerical resolution), will tend to select gas with higher densities and higher temperatures if the metallicity is lower. This can be seen from Fig.~\ref{fig:sfcriteria} by comparing the intersections of an instability line with the thermal equilibrium lines for different metallicities. Motivated by this observation, in Fig.~\ref{fig:Z-temps} we show the cumulative contributions of different densities (left) and temperatures (right) to the total SFR for different metallicities. For high metallicities ($Z\ge 0.1 Z_\odot$) the SFR is dominated by cold gas ($T\lesssim 10^2\,$K), while for very low metallicities ($Z\le 10^{-2} Z_\odot$) the SFR is dominated by warm ($T\gtrsim 10^3\,$K) gas. We emphasize that while the trend with metallicity is generic, the fractions of stars forming at different densities and temperatures depend explicitly on the numerical resolution, because we evaluate the gravitational instability criterion at the minimum resolved mass. An increase in numerical resolution leads to a smaller mass that has to become unstable, which requires higher densities and/or lower temperatures (see Fig.~\ref{fig:sfcriteria}). 

Fig.~\ref{fig:KS-importance-Z} shows the metallicity dependence of the spatially-resolved neutral (left) and molecular (right) KS relations. Unlike Fig.~\ref{fig:Z-temps}, these predictions are for observables and do not depend explicitly on the numerical resolution, because we average over a fixed spatial scale (750~pc) that is not explicitly tied to the resolution (which does not automatically mean the results are converged with the resolution, but they at least could be). Decreasing the metallicity from solar to primordial shifts the break in the neutral KS relation to around one order of magnitude higher neutral surface densities, but has little effect at higher surface densities (in agreement with simulations with a different SF criterion, e.g. \citealt{richings2016}).  The molecular KS relation is affected more strongly, it shifts to lower surface densities and has a much higher $\Sigma_{\rm SFR}$ at fixed $\Sigma_{\rm H_2}$. We caution, however, that at low surface densities the molecular KS law is sensitive to the resolution (see Fig.~\ref{fig:KS-resolution}).  
The difference in the molecular KS relation is dominated by the metallicity dependence of the molecular fraction, at lower abundances and a constant density there is less dust to catalyse molecule formation and shield dissociating radiation. In Appendix~\ref{app:temp-metals}, we show the neutral and molecular KS relations for the temperature ceiling SF criterion. Simulations using that criterion give a similar dependence on the metallicity although the break in the neutral KS relation is more sensitive to the metallicity than if the gravitational instability criterion is used. This is expected because while for the fiducial criterion a lower metallicity may result in the selection of higher density but warmer gas (see Fig.~\ref{fig:sfcriteria}), for the temperature ceiling criterion the gas must be cold and in thermal equilibrium, meaning that its density must increase further. In the next section we will study the impact of the SF criterion, but for the case of solar metallicity. 

%% file: subgrid_variations.tex
\subsection{The impact of subgrid physics variations} \label{subsec:subgrid_variations}

In this section, we focus on the impact of changing subgrid physics parameters on the spatially-resolved $t_\text{gas}$ relations. In particular, we will compare variations in the SF criterion (\S\ref{subsubsec:criterion}), the SF efficiency (\S\ref{subsubsec:SFE}), and the SN feedback energy (\S\ref{subsubsec:ESN}). Table \ref{tbl:subgridvar} lists all the different simulation variations. 

\begin{table}
\begin{center}
\caption{Variations in the subgrid parameters studied in this work. The columns list, from left to right, the simulation identifier,
  the adopted SF criterion (SF crit.),
  the energy per core collapse SN ($E_{\rm CC SN}$), 
  the SF efficiency per free fall time in the Schmidt law ($\varepsilon$). Parameter values that differ from the fiducial model are shown in bold face.
  All simulations use our fiducial particle mass ($10^5~\rm M_\odot$), solar metallicity and a kinetic feedback fraction of $f_{\rm kin}=0.1$.
\label{tbl:subgridvar}}
\begin{tabular}{lllrl}

\hline
  Simulation & SF crit. & $E_{\rm CC SN}$ & $\varepsilon$  \\
             & & ($10^{51} \rm erg$) & \\
\hline
  \texttt{fid}       & $\alpha_{\rm crit} = 1$ & $2.0$ & $0.01$  \\
  \texttt{alphacrit01}       & $\mathbf{\boldsymbol\alpha_{\rm crit} = 0.1}$ & $2.0$ & $0.01$\\
  \texttt{alphacrit03}       & $\mathbf{\boldsymbol\alpha_{\rm crit} = 0.3}$ & $2.0$ & $0.01$\\
  \texttt{alphacrit3}       & $\mathbf{\boldsymbol\alpha_{\rm crit} = 3}$ & $2.0$ & $0.01$\\
  \texttt{alphacrit10}       & $\mathbf{\boldsymbol\alpha_{\rm crit} = 10}$ & $2.0$ & $0.01$ \\
  \texttt{alphacrit30}       & $\mathbf{\boldsymbol\alpha_{\rm crit} = 30}$ & $2.0$ & $0.01$ \\
  \texttt{T1e4K}       & $\mathbf{T_{\rm crit} = 10^4~\rm K}$ & $2.0$ & $0.01$  \\
  \texttt{T3e3K}       & $\mathbf{T_{\rm crit} = 3\times 10^3~\rm K}$ & $2.0$ & $0.01$  \\
  \texttt{T1e3K}       & $\mathbf{T_{\rm crit} = 10^3~\rm K}$ & $2.0$ & $0.01$  \\
  \texttt{T3e2K}       & $\mathbf{T_{\rm crit} = 3 \times 10^2~\rm K}$ & $2.0$ & $0.01$  \\
  \texttt{T1e2K}       & $\mathbf{T_{\rm crit} = 10^2~\rm K}$ & $2.0$ & $0.01$  \\
  \texttt{nH1e-2cm-3}       & $\mathbf{n_{\rm H, crit} = 0.01~\rm cm^{-3}}$ & $2.0$ & $0.01$ \\
  \texttt{nH1e-1cm-3}       & $\mathbf{n_{\rm H, crit} = 0.1~\rm cm^{-3}}$ & $2.0$ & $0.01$ \\
  \texttt{nH1e0cm-3}       & $\mathbf{n_{\rm H, crit} = 1~\rm cm^{-3}}$ & $2.0$ & $0.01$  \\
  \texttt{nH1e1cm-3}       & $\mathbf{n_{\rm H, crit} = 10~\rm cm^{-3}}$ & $2.0$ & $0.01$ \\
  \texttt{nH1e2cm-3}       & $\mathbf{n_{\rm H, crit} = 100~\rm cm^{-3}}$ & $2.0$ & $0.01$ \\
  \texttt{nH1e3cm-3}       & $\mathbf{n_{\rm H, crit} = 1000~\rm cm^{-3}}$ & $2.0$ & $0.01$ \\
  \texttt{SFE0.001}       & $\alpha_{\rm crit} = 1$ & $2.0$ & $\mathbf{0.001}$  \\
  \texttt{SFE0.1}       & $\alpha_{\rm crit} = 1$ & $2.0$ & $\mathbf{0.1}$  \\
  \texttt{SFE1.0}       & $\alpha_{\rm crit} = 1$ & $2.0$ & $\mathbf{1.0}$  \\
  \texttt{ECCSN0.25e51}       & $\alpha_{\rm crit} = 1$ & $\mathbf{0.25}$ & $0.01$ \\
  \texttt{ECCSN0.5e51}       & $\alpha_{\rm crit} = 1$ & $\mathbf{0.5}$ & $0.01$ \\
  \texttt{ECCSN1e51}       & $\alpha_{\rm crit} = 1$ & $\mathbf{1}$ & $0.01$ \\
  \texttt{ECCSN4e51}       & $\alpha_{\rm crit} = 1$ & $\mathbf{4}$ & $0.01$ \\
  \texttt{ECCSN8e51}       & $\alpha_{\rm crit} = 1$ & $\mathbf{8}$ & $0.01$ \\
\hline
\end{tabular}
\end{center}
\end{table}

\begin{figure}
    \includegraphics[width=\columnwidth]{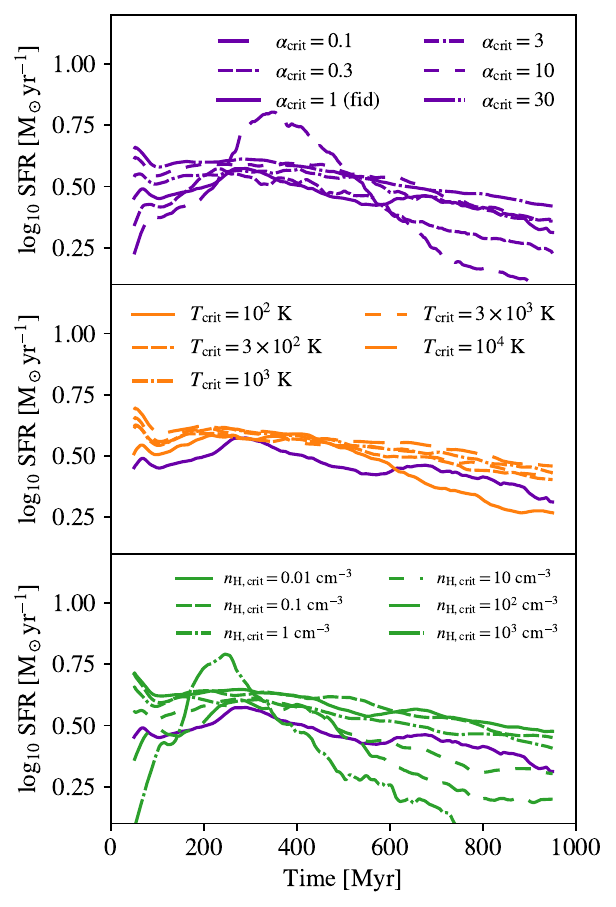} \vspace{-7mm}
    \caption{Comparison of the SFHs averaged over 100 Myr time scales for simulations using different SF criteria: a gravitational instability criterion (purple), a temperature ceiling (orange) or a density threshold (green). Different line styles correspond to different parameter values. The fiducial model is the solid purple curve, which is repeated in every panel. The SFHs are remarkably similar for most simulations. Only the three simulations with very strict SF criteria, i.e.\ $\alpha_{\rm crit}=0.1$, $n_{\rm H, crit} = 10^2~\rm cm^{-3}$, and $n_{\rm H, crit} = 10^3~\rm cm^{-3}$, predict a significantly different SFH.}
    \label{fig:SFH-crits}
\end{figure}

\begin{figure*}
    \includegraphics[width=\linewidth]{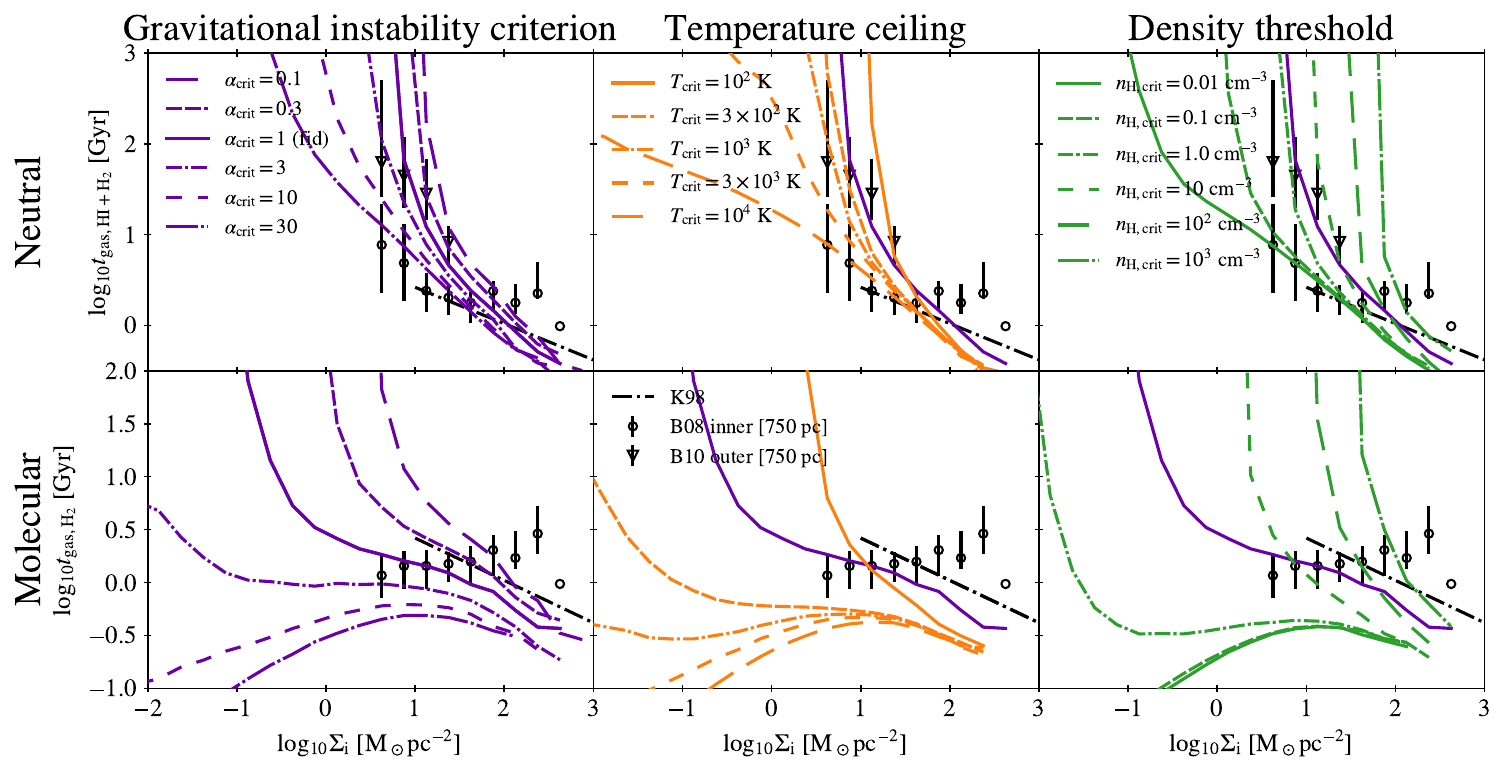} \vspace{-5mm}
    \caption{Comparison of the spatially resolved (750~pc) neutral (top) and molecular (bottom) median $t_{\rm gas}$ relations for different SF criteria. The left, middle, and right panels show parameter variations for, respectively, the gravitational instability criterion, a temperature ceiling, and a density threshold. The solid purples line indicates the fiducial model, $\alpha_\text{crit}<1$, which is repeated in every panel. Comparison with the data rules out gravitational instability criteria with $\alpha_\text{crit} \ll 1$ or $\alpha_\text{crit} \gg 1$ and also all the models with a temperature ceiling or density threshold.}
    \label{fig:KS-crits}
\end{figure*}

\subsubsection{The star formation criterion} \label{subsubsec:criterion}
Here we show the impact of the different SF criteria discussed in Section~\ref{sec:starformation} on the $t_{\rm gas}$ relations. We compare different normalisations of the gravitational instability criterion, different temperature ceilings and different density thresholds.

Fig. \ref{fig:SFH-crits} shows the SFHs for the simulations using different SF criteria. The SFHs have been averaged over 100 Myr to remove the short-time scale scatter. All simulations use our fiducial resolution ($M_\text{particle} = 10^5\,{\rm M}_\odot)$ and solar metallicity. Despite the huge differences in SF criteria, the SFHs are remarkably similar. Only the most restrictive criteria, i.e.\ $\alpha_\text{crit} \leq 0.1$ and $n_{\rm H, crit} \geq 100~\rm cm^{-3}$ result in significantly different SFHs. These simulations predict lower SFRs, probably because the numerical resolution becomes a bottleneck for reaching the densities required for SF. Relaxing the criteria a lot does not produce a significant change in the SFH, which indicates that the SFH is determined mostly by self-regulation through stellar feedback. 

Fig. \ref{fig:KS-crits} shows the spatially-resolved neutral (top) and molecular (bottom) $t_{\rm gas}$ relations for the different SF criteria. As expected, the results diverge towards low surface densities. There are nevertheless large differences between the models in the surface density range for which there are observational constraints. The most relaxed and the most strict gravitational instability criteria predict, respectively, too short and too long gas consumption time scales, particularly for molecular gas. Only the fiducial $\alpha_\text{crit} = 1$ and the $\alpha_\text{crit} = 3$ models appear consistent with the data. Only the lowest temperature ceiling ($T_{\rm crit} = 10^2\,$K) agrees with the observed molecular $t_\text{gas}$ and the data rule out all density threshold criteria. Many of the simulations predict $t_\text{gas}$ relations with the wrong shape, which suggests that for most SF criteria tuning the SFE parameter would not resolve the discrepancy with the data (for $T_{\rm crit} = 3 \times 10^2 ~\rm K$ or $10^3~\rm K$ the SFE parameter could be tuned).  

Overall, we find that in order to choose between subgrid SF criteria we should not focus on the SFH as this differs very little for vastly different criteria. This implies that the galaxy-averaged KS relation is also not suitable, at least not for idealized, isolated galaxies.
Furthermore, we should not choose based on properties of the gas evaluated at the resolution scale, like stellar birth densities and temperatures, because these are typically not converged and are not measurable. Instead, we find that a suitable way to determine whether SF criteria are appropriate is to compare the spatially-resolved $t_\text{gas}$ (or KS) relations.

\begin{figure}
    \includegraphics[width=\columnwidth]{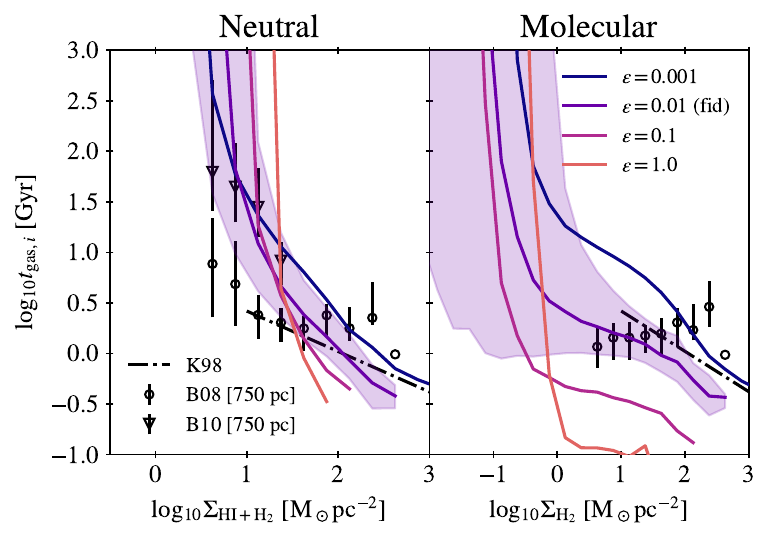}\vspace{-7mm}
    \caption{Comparison of the spatially-resolved (750 pc) neutral (left) and molecular (right) median $t_{\rm gas}$ relations for different star formation efficiencies (SFEs) per free fall time, $\varepsilon$ (different colours). The shaded regions show the 16\textsuperscript{th} and 84\textsuperscript{th} percentile scatter for the fiducial value of $\varepsilon=0.01$. For reference, the disk-averaged $t_{\rm gas}$ relation from \citetalias{kennicutt1998} (dot-dashed line) and the spatially-resolved $t_{\rm gas}$ relations from \citetalias{bigiel2008} and \citetalias{bigiel2010} are shown. At high surface densities, where $t_\text{gas}\ll 10$~Gyr, the gas consumption time scale decreases with the SFE, but the dependence is sublinear, particularly for neutral gas.}
    \label{fig:sfeff}
\end{figure}

\subsubsection{The star formation efficiency} \label{subsubsec:SFE}
The SFH averaged on 100 Myr time scales is not strongly affected by the SFE, except for the initial $\sim 100$~Myr (not shown). Because the total SFR is relatively unaffected by the choice of SFE, it is not a suitable observable for the calibration of $\varepsilon$. As was the case for the SF criterion,  the KS relations, particularly the spatially resolved, molecular ones, are however sensitive to the value of $\varepsilon$. 

Fig. \ref{fig:sfeff} shows the impact that the SFE, $\varepsilon$, has on the neutral (left) and molecular (right) $t_{\rm gas}$ relations. Besides the fiducial simulation with $\varepsilon=0.01$, we show simulations with $\varepsilon=0.001$, 0.1 and 1. For high SFEs the gas consumption time scales increase very steeply at low surface densities, creating a break in the $t_\text{gas}$ relations. At higher surface densities the trend is as expected: a higher SFE results in a shorter gas consumption time scale. However, at fixed surface density the increase of $t_{\rm gas}$ with $\varepsilon$ is sublinear. Increasing the SFE by a factor of 1000 only results in a decrease of $t_{\rm gas,H_2}$ by a factor $\sim 100$, and for neutral gas the change in the normalisation of the $t_\text{gas}$ relation is even smaller. The break in the neutral $t_{\rm gas}$ relation shifts to higher surface densities when the SFE is increased. The change in the break of the neutral $t_{\rm gas}$ relation is caused by the slightly increased velocity dispersion and temperature in the simulations with higher SFE, likely because SN feedback is more efficient when stars form more rapidly and at lower densities. 

\begin{figure}
    \includegraphics[width=\columnwidth]{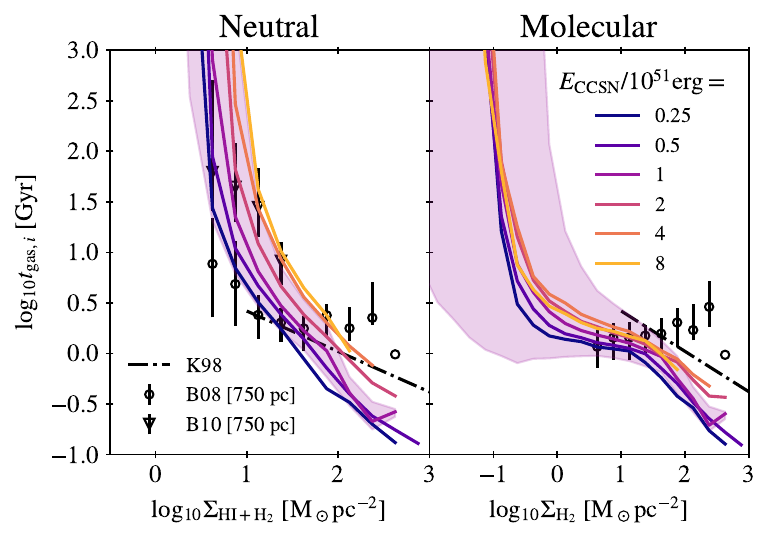} \vspace{-7mm}
    \caption{Comparison of the spatially resolved (750 pc) neutral (left) and molecular (right) median $t_{\rm gas}$ relations for different energies per core collapse supernova (different colours). The shaded regions show the 16\textsuperscript{th} and 84\textsuperscript{th} percentile scatter for the fiducial value of $E_\text{CCSN} = 2.5\times 10^{51}\,$erg. For reference, the disk-averaged $t_{\rm gas}$ relation from \citetalias{kennicutt1998} (dot-dashed line) and the spatially-resolved $t_{\rm gas}$ relations from \citetalias{bigiel2008} and \citetalias{bigiel2010} are shown.
    The gas consumption time scale increases with the feedback energy, but the dependence is weak except at very low surface densities. At high surface densities the $t_\text{gas}$ relations, particularly the molecular one, converge with increasing $E_\text{CCSN}$.}
    \label{fig:subgridESNII}
\end{figure}

\subsubsection{Supernova feedback} \label{subsubsec:ESN}
Fig. \ref{fig:subgridESNII} shows the neutral (left) and molecular (right) spatially-resolved $t_{\rm gas}$ relation for simulations with different values of $E_{\rm CC SN}$. The neutral $t_{\rm gas}$ relation shifts to about half a dex higher surface densities when $E_{\rm CC SN}$ is increased by a factor of 32 ($\approx 1.5$ dex). Because the relation steepens towards low surface densities, a constant shift implies that the increase in the neutral gas consumption is smaller at higher surface densities. At $\Sigma_{\ion{H}{I} + \rm H_2}\sim \rm 10^2\,M_\odot\,pc^{-2}$ $t_\text{gas}$ increases by about 0.5~dex for an 1.5~dex increase in $E_{\rm CC SN}$. The molecular gas consumption time scale also increases with $E_{\rm CC SN}$, and more so at lower surface densities, but the dependence is weaker than for the neutral gas. The two highest feedback energies ($E_{\rm CC SN}/10^{51}\,\text{erg} = 4$ and 8) give nearly identical results. Values of $E_{\rm CC SN} \sim 1-2\times 10^{51}\,$erg are most consistent with the data.

The galaxy SF histories (not shown) are much more sensitive to the feedback energy than to the SF parameters. As we have shown, the opposite is the case for the spatially resolved $t_\text{gas}$ (or KS) relations. The latter are therefore a better calibration target for the SF model, while the SFRs (or, for cosmological initial conditions, the stellar masses) can be used to calibrate the CC SN feedback parameters. However, the $t_\text{gas}$ relations are not independent of the feedback energy. Furthermore, other subgrid feedback parameters, like the fraction of energy injected in kinetic form, also influence the spatially-resolved KS laws \citep{chaikin2022b}. This means that care needs to be taken when calibrating subgrid models and that it may be necessary to recalibrate the SF model if the feedback prescription is modified.

%% file: softening.tex
\subsection{The impact of gravitational softening} \label{subsec:softening}

To avoid unrealistic two-body scattering, the gravitational force between particles in our simulations is reduced for separations smaller than a fixed gravitational softening length, $\epsilon$. This means that above a certain density, where the separation between particles is smaller than the softening length, we do not correctly simulate the collapse of clouds.
The density above which we begin to underestimate the self-gravity of a gas cloud is
\begin{align}
    n_{\rm H, soft} &= \frac{4\pi}{3}\frac{M_{\rm particle}}{\epsilon^3} \frac{X}{m_{\rm H}}, \label{eq:softening-rho} \\
    &\approx 16 {~\rm cm^{-3}} \left( \frac{M_{\rm particle}}{10^5 ~\rm M_\odot} \right) \left( \frac{\epsilon}{100~\rm pc} \right)^{-3},
\end{align}
where we use $X=0.756$. For clouds with higher densities self-gravity is not Newtonian (i.e. $F$ $\propto$ $r^{-2}$) while the gravitational instability criterion presented in \S\ref{subsub:instability} is based on the Newtonian \citet{jeans1902} mass. In Ploeckinger et al. (in prep) we introduced the softened Jeans criteria which extends the \citet{jeans1902} to be applicable in the case of softened gravitational forces. In Ploeckinger et al. (in prep) we studied gravitational (in)stabilities at the resolution limit of Lagrangian simulations assuming only thermal support (i.e. pressure). Here we extend this analysis by including turbulent support and applying the softened \citet{jeans1902} criterion to the SF criterion.

\begin{figure*}
    \centering
    \includegraphics[width=.9\textwidth]{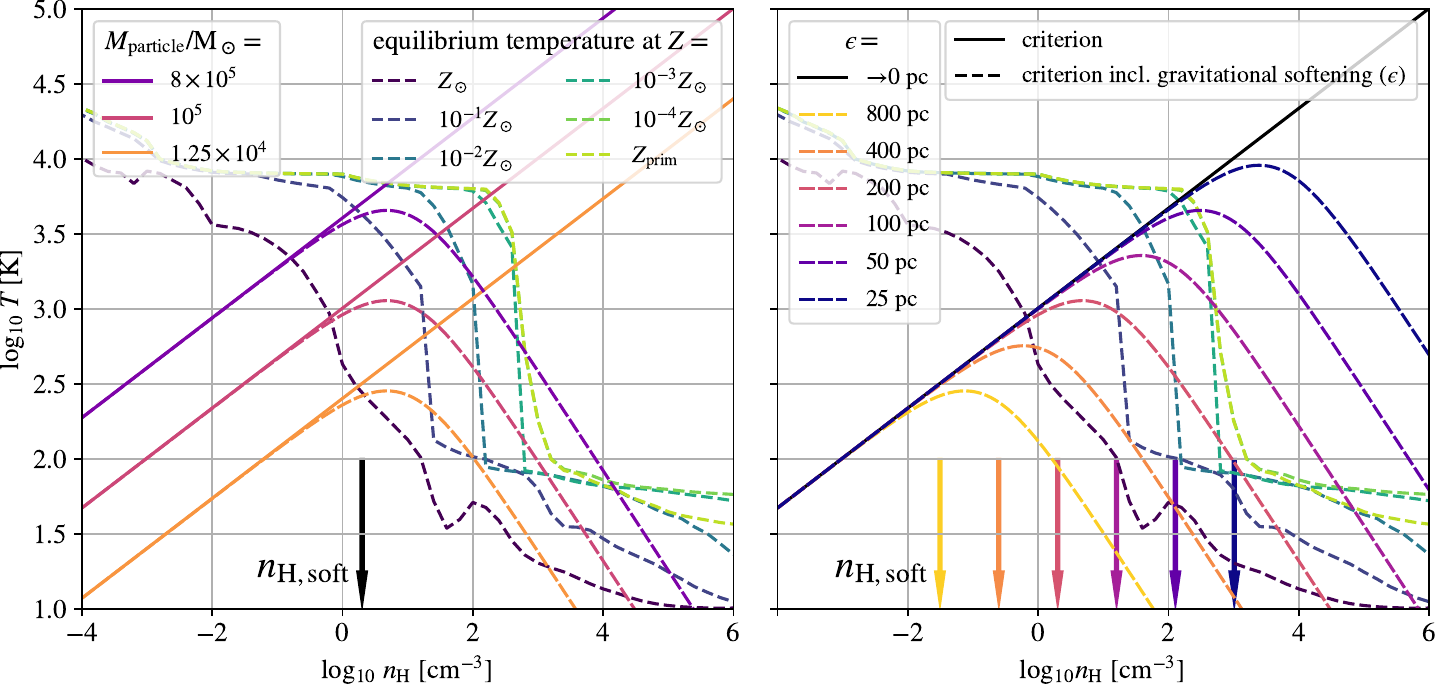}
    \caption{The gravitational instability criterion ($\alpha_{\rm crit}=1$) in temperature-density space for softened (dashed lines) and unsoftened (solid lines) gravitational forces assuming $\sigma_\text{th}\gg \sigma_{\rm turb}$. The left panel compares different particle masses (different colours) while keeping the ratio $M_\text{particle}/\epsilon^3$ fixed and $\alpha_{\rm crit}=1$. The right panel compares different gravitational softening lengths (different colours) at our fiducial particle mass of $10^5~\rm M_\odot$ and $\alpha_{\rm crit}=1$. For reference, we plot the thermal equilibrium temperature curves for different metallicities (dotted lines, different colours) for the  Ploeckinger et al. (in prep.) cooling tables. The arrows indicate $n_{\rm H, soft}$, the density above which self-gravity begins to be softened for our fixed ratio of $M_\text{particle}/\epsilon^3$ (left) and for different gravitational softening lengths (right). Gas clouds with the minimum resolved mass, $\langle N_{\rm ngb} \rangle M_\text{particle}$, with physical conditions falling in the regions below a given line are gravitationally unstable. In between the solid and dashed lines, such gas clouds should be unstable, but softening keeps them stable. A smaller gravitational softening length moves the softened gravitational instability curve to higher densities.
    }
    \label{fig:sfcriteria-res}
\end{figure*}

To investigate the direct impact of softening on the SF criterion, we modify the derivation of the gravitational instability criterion from \S\ref{subsub:instability} such that it is valid for the case of softened gravity in the limit $\epsilon \gg r$. While the sound and turbulence crossing times across a cloud with radius $h$ and mass $m$ remain identical, the free-fall time of the collapsing cloud in \citeauthor{plummer1911}-softened gravity (i.e. $\Phi \propto 1/\sqrt{r^2 + \epsilon^2}$) becomes
\begin{align}
 t_{\rm ff,soft} &= -\frac{1}{\sqrt{2Gm}}\int\limits_h^0 \left( \frac{1}{\sqrt{r^2 + \epsilon^2}} - \frac{1}{\sqrt{h^2 + \epsilon^2}} \right)^{-1/2} {~\rm d}r, \\
 &= t_{\rm ff} \sqrt{ 1 + 2\left( \frac{\epsilon}{h} \right)^3 }, \label{eq:soft-ff-time}
\end{align}
which is the analog of equation (\ref{eq:velocity}). In the limit of $\epsilon \gg h$, we obtain
\begin{align}
    t_{\rm ff,soft} &\approx \sqrt{2} t_{\rm ff} \left( \frac{\epsilon}{h} \right)^{3/2 }, \\
    &\approx 64~{\rm Myr} \left( \frac{n_{\rm H}}{1 ~\rm cm^{-3}} \right)^{-1/2} \left( \frac{\epsilon}{100~\rm pc} \right)^{3/2} \left( \frac{h}{100~\rm pc} \right)^{-3/2}.
\end{align}
Therefore, the gas cloud is gravitationally unstable when
\begin{align}
    \sqrt{2} t_{\rm ff} \left( \frac{\epsilon}{h} \right)^{3/2 } < \frac{h}{\sqrt{\sigma_{\rm th}^2 + \sigma^2_{\rm turb}}}.
\end{align}
Setting $h = \lambda_{\rm J,soft}$ gives the \citet{jeans1902} length in the limit $\epsilon\gg h$,
\begin{align}
    \lambda_{\rm J, soft} &= \left( \frac{6\pi}{32 G \rho} \right)^{1/5} \epsilon^{3/5} \left( \sigma_{\rm T}^2 + \sigma_{\rm turb}^2 \right)^{1/5}, \\
    &\approx 3.2\times 10^2\,{\rm pc} \left( \frac{n_{\rm H}}{1 ~\rm cm^{-3}} \right)^{-1/5} \left( \frac{\epsilon}{200~\rm pc} \right)^{3/5} \left( \frac{\sqrt{\sigma_{\rm T}^2 + \sigma_{\rm turb}^2}}{10~\rm km ~\rm s^{-1}} \right)^{2/5},
\end{align}
with a corresponding \citet{jeans1902} mass of
\begin{align}
    M_{\rm J,soft} &= \frac{4\pi}{3} \rho^{2/5} \left( \frac{6\pi}{32 G} \right)^{3/5} \epsilon^{9/5} \left( \sigma_{\rm T}^2 + \sigma_{\rm turb}^2 \right)^{3/5}, \\
    &\approx 4.5 \times 10^6 ~{\rm M_\odot} \left( \frac{n_{\rm H}}{1 ~\rm cm^{-3}} \right)^{2/5} \left( \frac{\epsilon}{200~\rm pc} \right)^{9/5} \left( \frac{\sqrt{\sigma_{\rm T}^2 + \sigma_{\rm turb}^2}}{10~\rm km ~\rm s^{-1}} \right)^{6/5}.
\end{align}
This implies that, at densities where self-gravity is strongly softened, the \citet{jeans1902} mass increases with the density and hence gas that is classified as unstable by the unsoftened instability criterion (discussed in \S \ref{subsub:instability}) may actually be stable. A gas cloud is gravitationally unstable when $m > M_{\rm J}$, where $m=\langle N_{\rm ngb} \rangle M_{\rm particle}$ is the mean mass in the kernel. Therefore, the analogue of equation (\ref{eq:instability-crit}) in the limit $h\ll \epsilon$ becomes
\begin{align}
    \alpha_{\rm soft} &\equiv \frac{\rho^{2/5} \epsilon^{9/5} (\sigma_{\rm T}^2 + \sigma_{\rm turb}^2)^{3/5}}{G^{3/5} \langle N_{\rm ngb} \rangle M_{\rm particle}} < \alpha_{\rm crit, soft},
\end{align}
\begin{align}
    \alpha_{\rm soft} &\equiv \frac{\rho^{2/5} (\sigma_{\rm T}^2 + \sigma_{\rm turb}^2)^{3/5}}{G^{3/5} \langle N_{\rm ngb} \rangle M_{\rm particle}^{2/5}} \left( \frac{\epsilon^3}{M_{\rm particle}}\right)^{3/5} < \alpha_{\rm crit, soft},
\end{align}
where $\alpha_{\rm crit, soft}$ is a constant of order unity given by $\alpha_{\rm crit, soft}=3 (32)^{3/5} /(4\pi (6\pi)^{3/5})\approx 0.33$ (for unsoftened gravity we had $\alpha_\text{crit}\approx 1.3)$.
In the limiting case of $\sigma_{\rm th} \gg \sigma_{\rm turb}$ the instability criterion becomes an upper limit on the temperature,
\begin{align}
    T <& \; 9.7 \times 10^3 ~{\rm K} \left( \frac{n_{\rm H}}{1 ~\rm cm^{-3}} \right)^{-2/3}  \left( \frac{\langle N_{\rm ngb} \rangle}{65} \right)^{5/3} \times \label{eq:softened-T} \\
    &\;\left( \frac{M_{\rm particle}}{10^5~\rm M_\odot} \right)^{5/3} \alpha_{\rm crit, soft}^{5/3} \left( \frac{\epsilon}{200~\rm pc} \right)^{-3} , \notag \\
\end{align}
or
\begin{align}
   T <& \; 9.7 \times 10^3 ~{\rm K} \left( \frac{n_{\rm H}}{1 ~\rm cm^{-3}} \right)^{-2/3}  \left( \frac{\langle N_{\rm ngb} \rangle}{65} \right)^{5/3} \times \label{eq:softened-T2} \\
    &\;\left( \frac{M_{\rm particle}}{10^5~\rm M_\odot} \right)^{2/3} \alpha_{\rm crit, soft}^{5/3} \left( \frac{M_{\rm particle}/\epsilon^3}{10^5~\rm M_\odot/(200~\rm pc)^3} \right)\ , \notag
\end{align}
where we assumed a mean particle mass $\mu=1.3$ and a hydrogen mass fraction $X = 0.756$. Note that instability requires a lower temperature if the softening length is larger.

The limiting case of $\sigma_{\rm turb} \gg \sigma_{\rm th}$ gives an upper limit on the density of
\begin{align}
    n_{\rm H} 
    <& \; 7.4 \times 10^3 ~{\rm cm^{-3}} \left( \frac{\sigma_{\rm turb}}{5~\rm km ~\rm s^{-1}} \right)^{-3} \left( \frac{\langle N_{\rm ngb} \rangle}{65} \right)^{5/2} \times\\ &\;
     \left( \frac{M_{\rm particle}}{10^5~\rm M_\odot} \right)^{5/2} \alpha_{\rm crit, soft}^{5/2} \left( \frac{\epsilon}{100~\rm pc} \right)^{-9/2} \ , \notag \\
\end{align}
or
\begin{align}
     n_{\rm H} <& \; 7.4 \times 10^3 ~{\rm cm^{-3}} \left( \frac{\sigma_{\rm turb}}{5~\rm km ~\rm s^{-1}} \right)^{-3} \left( \frac{\langle N_{\rm ngb} \rangle}{65} \right)^{5/2} \times\\ &\;
     \left( \frac{M_{\rm particle}}{10^5~\rm M_\odot} \right) \alpha_{\rm crit, soft}^{5/2} \left( \frac{M_{\rm particle}/\epsilon^3}{10^5~\rm M_\odot/(200~\rm pc)^3} \right)^{3/2} \ . \notag
\end{align}
This means that to resolve instabilities at higher densities the softening length is required to be smaller.

\begin{figure}
    \includegraphics[width=\columnwidth]{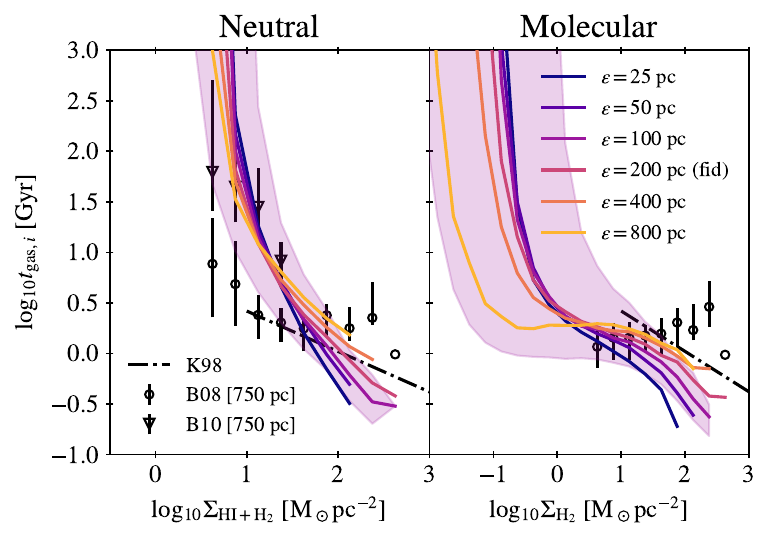} \vspace{-7mm}
    \caption{
    Comparison of the spatially resolved (750 pc) neutral (left) and molecular (right) median $t_{\rm gas}$ relations for different gravitational softening lengths (different colours). The shaded regions show the 16\textsuperscript{th} and 84\textsuperscript{th} percentile scatter for the fiducial value $\epsilon = 100~\rm pc$. For reference, the disk-averaged $t_{\rm gas}$ relation from \citetalias{kennicutt1998} (dot-dashed line) and the spatially-resolved $t_{\rm gas}$ relations from \citetalias{bigiel2008} and \citetalias{bigiel2010} are shown. At low surface densities the gas consumption time scale is almost independent of the gravitational softening length as long as it is not too large ($\epsilon \lesssim 400 ~\rm pc$). At high surface densities the gas consumption time scale decreases systematically with increasing force resolution, though the dependence is weak.}
    \label{fig:KS-grav-sof}
\end{figure}

The left panel of Fig.~\ref{fig:sfcriteria-res} shows the gravitational instability criterion derived in \S\ref{subsub:instability} in temperature-density space (solid lines) for different values of the particle mass $M_\text{particle}$ (different colours) and $\alpha_\text{crit}=1$. This is compared with the softened gravitational instability criterion in the limit $\sigma_\text{th}\gg \sigma_{\rm turb}$ and assuming $\epsilon\propto M_\text{particle}^{1/3}$ (dashed lines). For the softened gravitational instability criterion we derived the limit of $\epsilon \gg h$ and interpolate between this regime and the normal gravitational instability criterion using $T_{\epsilon} = (T_{h \gg \epsilon}^{-1} + T_{\epsilon \gg h}^{-1} )^{-1}$. This ensures a smooth transition between the two limiting cases. The left panel of Fig.~\ref{fig:sfcriteria-res} shows that the density above which the softened instability criterion starts to deviate from the unsoftened criterion corresponds to $n_{\rm H,soft}$ as expected.

Gas below the dashed curves in the left panel of Fig.~\ref{fig:sfcriteria-res} is gravitationally unstable\footnote{However, it has a longer free-fall time than the Newtonian free-fall time, so the collapse takes longer and follows equation (\ref{eq:soft-ff-time}).}, while gas between the solid and dashed curves should be unstable, but is stabilized by gravitational softening. The thermal equilibrium temperatures (dotted lines) are located mostly below the dashed curves, which suggests that in practice gravitational softening will generally not directly prevent SF (assuming $\sigma_\text{th}\gg \sigma_{\rm turb}$). For our fiducial resolution, $M_\text{particle}=10^5\,\text{M}_\odot$, the equilibrium temperature of metal-free gas does extend in the problematic regime between the solid and dashed curves and the same is true for low-metallicity gas at very high densities, $n_\text{H} > 10^{4}~\text{cm}^{-3}$. However, these problematic densities exceed  $n_{\rm H, soft}$ by many orders of magnitude, which means self-gravity is unlikely to produce them, though they may occur in pressure-confined clouds and in the centres of clouds with mass $\gg \langle N_{\rm ngb} \rangle M_\text{part}$ owing to the weight of the overlying layers.

For fixed $M_\text{particle}/\epsilon^3$, lower-resolution simulations, i.e.\ simulations using a larger $M_\text{particle}$, are able to resolve gravitational instabilities up to higher densities (i.e., the dashed instability curves in the left panel of Fig.~\ref{fig:sfcriteria-res} shift to higher densities for a fixed temperature). This may sound counterintuitive, but it is a consequence of the fact that we evaluate the instability criterion at the resolution limit, i.e.\ for a cloud mass that scales with the particle mass. This suggests that keeping $M_{\rm particle}/\epsilon^3$ fixed may not be desirable when the particle mass is decreased to small values.

In the right panel of Fig.~\ref{fig:sfcriteria-res} we again show the unsoftened and softened gravitational instability criteria in temperature-density space for our fiducial resolution and assuming $\sigma_\text{th}\gg \sigma_{\rm turb}$ and compare it with gravitational instability criteria for smaller gravitational softening lengths, keeping the particle mass fixed at our fiducial value. This shows that when the gravitational softening length is a factor of two smaller, the softened gravitational instability curve begins to deviate from the line corresponding to the unsoftened gravitational softening length at a factor eight higher densities. This is the opposite trend as was found by Ploeckinger et al. in prep. for artificial collapse triggered by the minimum SPH smoothing length. However, the impact at high densities is larger because equation (\ref{eq:softened-T}) scales strongly with $\epsilon$ and we obtain a softening dependent density at which $10~\rm K$ gas is unstable given by
\begin{align}
    n_{\rm H, \epsilon} &< 3.0 \times 10^{\:\!4} {~\rm cm^{-3}} \left (\frac{T}{10~\text{K}}\right )^{-3/2} \left( \frac{\epsilon}{200~\rm pc} \right)^{-9/2}. \label{eq:limit-nH}
\end{align}
This means that a factor of two in the gravitational softening length at fixed particle mass will change the maximal density for which the gas is unstable by 1.35~dex.

Fig. \ref{fig:KS-grav-sof} shows the $t_{\rm gas}$ relation for simulations with different gravitational softening lengths. The $t_{\rm gas}$ relations depend only weakly on the gravitational softening lengths. The break in the relations shifts to smaller molecular surface densities for the largest gravitational softening length ($800 ~\rm pc$). At high surface densities the gas consumption time scales increase systematically with the softening length, but the effect is small compared with the scatter.

Altogether we find that a softened or unsoftened gravitational instability criterion selects almost the same gas when the softening length is small enough for the mass resolution. The gas considered stable in softened gravity is mainly at high densities and higher temperatures than the equilibrium temperatures. Because of this, the impact on the star-forming gas is limited and the KS relations are not influenced much.

If we want to resolve the same region in the $\rho-T$ plane, we can use equation (\ref{eq:softened-T}) to give a relation between the particle resolution and the gravitational softening length of
\begin{align}
    \epsilon &= 200~{\rm pc} \left( \frac{M_{\rm particle}}{10^5~\rm M_\odot} \right)^{5/9},
\end{align}
which has a stronger dependence on the particle mass as compared to the typical scaling (i.e. $\epsilon \propto M_{\rm particle}^{1/3}$). However, this still does not allow us to resolve higher density than the limit given by equation (\ref{eq:limit-nH}). If we use a scaling of the gravitational softening length given by
\begin{align}
    \epsilon &= 200~{\rm pc} \left( \frac{M_{\rm particle}}{10^5~\rm M_\odot} \right)^{n},
\end{align}
then it is required that $n>5/9$ to resolve higher densities correctly for higher resolutions. A slight increase gives a value of $n=2/3$. This would always resolve the same region of $\rho-T$ and would allow us to resolve higher densities when we have higher resolution. However, using similar small gravitational softening lengths for stellar particles could result in undesired artificial scattering of stellar particles \citep{ludlow2019,ludlow2020,ludlow2021,wilkinson2022}.

%% file: resolution.tex
\subsection{Comparison with the literature}
This is not the first attempt to understand the SF scaling relations in hydrodynamical simulations. However, many impose an equation of state to prevent the gas from cooling down to low ($T\ll10^{\:\!4}~\rm K$) temperatures \citep[e.g.][]{springel2003,stinson2006,schaye2008}. \citet{marinacci2019} showed that there are significant differences between hydrodynamical simulations with and without an equation of state, such as a stronger variation of the SFH on time scales of $t\sim 20~\rm Myr$ and a larger scatter in $\Sigma_{\rm SFR}$ at fixed gas surface densities. We therefore focus on comparing our results with simulations that explicitly model cold gas ($T\lesssim 10^{\:\!2} ~\rm K$).

Simulations of idealised disk galaxies indicate that the SFH is insensitive to the SF criterion. This is the case for a wide variation of SF criteria, such as density thresholds, temperature ceilings, converging flows, $t_{\rm cool}<t_{\rm ff}$, gravitational instability and combinations thereof (e.g. \citealt{hopkins2013}; \citealt{sillero2021}; and our Fig.~\ref{fig:SFH-crits}) and it is also true for cosmological simulations \citep[e.g.][]{hopkins2018,hopkins2022} and cosmological simulations that use an equation of state \citep{schaye2010}. The SF criterion does however have a significant effect on the physical conditions of the gas from which stars form \citep[e.g.][]{hopkins2013,hopkins2022} and very strict criteria, like density thresholds that are high relative to the numerical resolution, can produce artificial features \citep[e.g.\ Fig.~6 of][]{hopkins2022}. Furthermore, the combination of multiple SF criteria is dominated by the most strict SF criterion (e.g. a density threshold in FIRE2 and a gravitational instability criterion in FIRE3, see Figs. 6 and A1 of \citealt{hopkins2022}). \citet{becerra2014} found that with higher density thresholds the normalisation of the total $t_{\rm gas}$ relation decreases. This is opposite to what we are finding (see Fig.~\ref{fig:KS-crits}). While a higher density threshold leads to stars forming at higher surface densities and hence from gas with shorter consumption time scales, the fraction of gas that is star forming decreases, which leads to longer gas consumption time scales. Apparently, in our simulations the latter effect wins, yielding higher normalisations of the $t_{\rm gas}$ relations for higher density thresholds.

Matching the observed high-surface density slope of the spatially-resolved and azimuthally-averaged neutral gas KS relation ($N_{\rm KS}\approx 1.4$) or, equivalently, the slope of the $t_{\rm gas}$ relation ($N\approx -0.4$) is challenging. Some simulations agree with the canonical values \citep[e.g.][]{becerra2014,semenov2016,semenov2017, sillero2021}, while we ($N\approx -0.7$, see Fig. \ref{fig:KS-neutral}) and others predict relations that are slightly steeper \citep[$N\approx -0.7$,][]{orr2018}. Other simulations predict relations that are so steep that they do not reproduce the observations ($N\approx -1.5$, \citealt{agertz2015}; $N\approx -3$ to $-5$, \citealt{marinacci2019}; $N\approx -1.5$, \citealt{gensior2020};  $N\approx -1$ to $ -4$, \citealt{bieri2022}).

The main ingredients responsible for the normalisation of the $t_\text{gas}$ relation are the SFE and the radiative cooling. Prescriptions for radiative cooling  that do not form enough molecular gas decrease the normalisation of the molecular $t_{\rm gas}$ relation (similar to the metallicity trend of Fig.~\ref{fig:KS-neutral}). This seems to be the case for \citet{semenov2017}, who found that their molecular $t_{\rm gas}$ relation has a 0.5~dex lower normalisation than observed, their neutral $t_{\rm gas}$ relation agrees with the data. The SFE does not change the average SFR significantly for values between 1 and 100 per cent \citep[e.g.][]{li2020}, but it can significantly change the normalisation of the $t_\text{gas}$ relation as shown in Fig.~\ref{fig:sfeff}. The FIRE2 simulations \citep{hopkins2018} use an SFE of 100 per cent, which results in a molecular $t_{\rm gas}$ relation with a 1.5~dex too low normalisation \citep[Fig.~2 of][]{orr2018}. \citet{agertz2013} showed that a 10 per cent SFE gives a too low normalisation while 1 per cent is in agreement with observations. Similarly, \citet{semenov2018} showed that a higher SFE decreases the normalisation significantly (their Fig.~10) and an SFE of 1 per cent is in agreement with the data. These results are all in agreement with what we find. This suggests that using a low SFE of $\sim 1$ per cent is required because we are still far away from resolving the densities at which actual stars form and the efficiency is close to 100 per cent.

\subsection{Which SF criterion is best?}
In \S \ref{subsubsec:criterion} we showed that the spatially-resolved $t_{\rm gas}$ relation is an excellent discriminator of SF criteria. Based on Fig.~\ref{fig:KS-crits} alone, we find that at the resolution studied in this paper and for solar metallicity, the gravitational instability criterion (for both $\alpha_{\rm crit}=1$ and $3$) and the temperature criterion (for $T_{\rm crit} \leq 10^{\:\!3}~\rm K$) all reproduce the observed $t_{\rm gas}$ relations (although for the temperature criterion you will need to adjust the SFE). However, as shown in Fig.~\ref{fig:KS-resolution}, a fixed criterion in $\rho$-$T$ space, which for gravitational instability at the resolution limit implies a value of $\alpha_\text{crit}$ that depends on the numerical resolution, results in poor convergence of the $t_{\rm gas}$ relation with the numerical resolution. Instead, a gravitational instability criterion with a fixed value of $\alpha_\text{crit}$, which corresponds to different regions in $\rho$-$T$ space for different numerical resolutions, reproduces the $t_{\rm gas}$ relation and converges much better with the numerical resolution.

For low-metallicity gas, the behaviour is significantly different for a temperature ceiling and a gravitational instability criterion. 
Very low metallicity gas requires very high densities to cool below temperatures of $T\approx 10^{\:\!3}~\rm K$, particularly if an interstellar radiation field is included. This means that cosmological simulations with a temperature criterion that do not resolve such high densities might be unable to form any stars at all. A gravitational instability criterion with $\alpha_\text{crit}\sim 1$ will however enable SF at higher temperatures in regions with sufficiently low turbulent velocity dispersion and sufficiently high densities.

While stars form in cold molecular clouds, generally not all cold gas is gravitationally unstable. Cold gas with a large turbulent dispersion will not be gravitationally unstable (see Fig.~\ref{fig:sfcriteria}) and hence should not be considered star forming. Therefore, situations that can cause an increase in the turbulent velocity, like the presence of a bulge or stellar feedback, will make it harder to form stars, e.g.\ leading to morphological quenching \citep[e.g.][]{martig2009,martig2013} or self-regulation. A temperature criterion that just determines if gas is cold is less sensitive to such effects (though supersonic turbulence may be converted to heat). 

Overall, these points indicate that a gravitational instability criterion (with a fixed $\alpha_\text{crit}\sim 1$) is the better choice for a SF criterion.

\subsection{Caveats}
The gas and stellar disks in our idealised disk galaxies are initialized with a constant density at a fixed radius. This means that any spiral structure in the gas is developed by internally triggered instabilities. However, spiral structure can develop due to interactions with satellite galaxies \citep[see][for a review]{dobbs2014}, a nearby galaxy cluster \citep[e.g.][]{semczuk2017}, asymmetries in the DM distribution \citep{khoperskov2013}, or tidal perturbations from DM subhaloes \citep[e.g.][]{chang2011}. \citet{pettitt2020} showed that including a satellite galaxy does impact the structure of the ISM at large radii ($r>3~\rm kpc$) producing higher densities in the arms and lower densities between them. Similarly, the initial gas fraction and metallicity are constant throughout the disk, which is unlikely to be true for real galaxies. 

Furthermore, our simulations lack a circumgalactic medium (CGM), the presence of cosmological accretion and a cosmological environment. This means outflows are able to escape with higher velocities and to larger distances than would be the case if a CGM were included and that there is no replenishment of gas consumed by SF. 
Therefore a limitation of idealised galaxy simulations is that they are probably unsuitable for investigating the spatially-resolved SF main sequence and gas main sequence.

Our adopted gas fraction of 30 per cent is high for galaxies observed at $z=0$. However, in order to test SF models, we need a high enough $\Sigma_{\rm \ion{H}{I}+H_2}$ to reach the power-law regime of the KS relations. A lower gas fraction would increase the stellar surface density at fixed $\Sigma_{\rm \ion{H}{I} + H_2}$, which may push the break in the neutral KS relation to slightly lower $\Sigma_{\rm \ion{H}{I}+H_2}$. Note that this would improve the agreement with the data.

Our disk galaxy does not contain a classical bulge component. Real galaxies and galaxies formed in cosmological simulations contain bulge components that vary in strength and are formed by the collisions of galaxies or disk instabilities \citep[see][for a review]{brooks2016}. The inclusion of a bulge component could suppress SF in the centre \citep[e.g.][]{martig2009,martig2013}. We leave an investigation of the effect of the bulge for future work.

Lastly, our simulations use a fixed gravitational softening length $\epsilon$, while an adaptive gravitational softening length is often used (e.g. \citealt{price2007} and models based on this like \citealt{springel2010,iannuzzi2011}). However, as we show in Fig. \ref{fig:KS-grav-sof} different gravitational softening lengths give almost identical results.

%% file: conclusion.tex
\section{Conclusions} \label{sec:conclusion}

We have used SPH simulations of isolated Milky Way mass disk galaxies that include cold, interstellar gas to test subgrid prescriptions for star formation (SF). Our fiducial model consists of a gravitational instability criterion that includes both thermal and turbulent motions and is evaluated at the mass resolution limit (equation~\ref{eq:instability-crit} with $\alpha_\text{crit}=1$) combined with the Schmidt SF law (equation~\ref{eq:schmidt-law}) with an efficiency per free fall time of $\varepsilon =0.01$. Our fiducial simulation assumes solar metallicity and uses a particle mass of $10^5\,\text{M}_\odot$ and a gravitational softening length of 200~pc. We investigated the impact of the numerical resolution, SF criterion, subgrid SF efficiency, supernova feedback energy, metallicity, and the binning of the observables on the predicted Kennicutt-Schmidt (KS) SF law for neutral, molecular and atomic gas. The main conclusions we draw are:
\begin{enumerate}
    \item Our fiducial model matches the observed neutral, molecular and atomic Kennicutt-Schmidt (KS) relations well for both spatially-resolved (Fig.~\ref{fig:KS-neutral}) and azimuthally-averaged data (Fig.~\ref{fig:radialvsspatial}). 
    \item Varying the spatial bin size between 100~pc and 10~kpc shows that the spatial averaging scale strongly affects the spatially-resolved KS relations. As expected, the scatter decreases with the spatial bin size. At low surface densities a smaller bin size results in a longer median gas consumption time scale. However, for a given bin size there is a surface density above which the neutral and molecular, but not the atomic, KS laws converge to the corresponding KS laws for larger bin sizes (Fig.~\ref{fig:spatialres}). Azimuthally-averaged KS laws using a bin size similar to the ones used in observations ($\sim 750$~pc) resemble spatially-resolved ones using larger bin sizes (similar to the disk scale length) (Fig.~\ref{fig:radialvsspatial}).  
    \item Decreasing the metallicity shifts the break in the neutral gas KS relation to somewhat higher surface densities. For the molecular KS law the effect is more dramatic, with lower metallicities yielding shorter gas consumption time scales at fixed molecular surface density because the molecular density more strongly underestimates the true gas density (Fig.~\ref{fig:KS-importance-Z}).      
    \item Observational and theoretical studies of SF laws generally ignore ionized gas. However, for neutral hydrogen surface densities $\ll 1~\rm M_\odot\,\rm pc^{-2}$ ionized gas dominates the total gas surface density and at a neutral surface density of $10~\rm M_\odot ~\rm pc^{-2}$ ionized gas still contributes about as much as molecular hydrogen does (for a spatial averaging scale of 750~pc; Fig.~\ref{fig:diff-species}).     
    \item To investigate the numerical convergence, we vary the mass (spatial) resolution by a factor of 4096 (16). We find good convergence for the SF history (Fig.~\ref{fig:SFHres}), for the neutral KS law, and at $\Sigma_{\text{H}_2}\gg 10\,\text{M}_\odot\, \text{pc}^{-2}$ also for the molecular KS law  (Fig.~\ref{fig:KS-resolution}). Even though we show that gravitational softening does modify the region of $\rho-T$ space that is unstable (Fig.~\ref{fig:sfcriteria-res}), we find that the KS laws are largely insensitive to the gravitational softening length (when varied between 50 and 800~pc at our fiducial mass resolution) although at very high surface densities ($\gtrsim 10^2\,\text{M}_\odot\, \text{pc}^{-2}$) the gas consumption time scale increases slightly with the softening length (Fig.~\ref{fig:KS-grav-sof}). 
    \item The gravitational instability criterion selects different regions of $T-n_{\rm H}$ plane depending on the mass resolution and the value of $\alpha_\text{crit}$. Higher (lower) mass resolution means that gas with higher (lower) densities and lower (higher) temperatures is selected to be star-forming (Fig.~\ref{fig:sfcriteria}). While this resolution dependence can largely be circumvented by scaling $\alpha_\text{crit}$ with the resolution (Fig.~\ref{fig:density-sf-gas}), we think it is desirable to evaluate the SF criterion at the simulation's resolution limit, because it allows higher-resolution models to directly simulate the physics up to higher densities. We thus argue that numerical convergence is desirable for observables, but not for the physical properties of gas that is labelled star-forming by the subgrid model. In fact, we find that a fixed value of $\alpha_\text{crit}$ gives better convergence for the KS laws than a value of $\alpha_\text{crit}$ that is scaled with the resolution (Fig.~\ref{fig:KS-resolution}).  
    \item We compare simulations using the gravitational instability criterion for different values of $\alpha_\text{crit}$ with simulations using a range of density thresholds or temperature ceilings. While SF histories are insensitive to the SF criterion (Fig.~\ref{fig:SFH-crits}), the KS relations can discriminate between different models (Fig.~\ref{fig:KS-crits}). Comparison with the data rules out gravitational instability criteria with $alpha_\text{crit} \ll 1$ or $\alpha_\text{crit} \gg 1$ and also all the models with a temperature ceiling or density threshold. Additionally, we prefer the instability criterion over a temperature ceiling because it is more robust for cosmological simulations, where (nearly) metal-free gas may only cool to low temperatures at very high densities that typically remain unresolved, and because physically we do not expect all cold gas to form stars. 
    \item Varying the SF efficiency per free-fall time by a factor of $10^3$ shows that it has little effect on the SF history, but has a strong impact on the normalization of and the break in the KS relations. Comparison with observations constrains the SF efficiency to be of order 1 per cent (Fig.~\ref{fig:sfeff}).
    \item Varying the strength of stellar feedback by a factor of 32 shows that it has a large impact on the SF history and a non-negligible effect on the normalization of the KS laws, particularly for the neutral gas (Fig.~\ref{fig:subgridESNII}).

In future work we plan to use our fiducial SF prescription in cosmological simulations of galaxy formation. This will enable us to include effects such as gas accretion and galaxy interactions, and to investigate a wide range of galaxy masses, gas fractions, and morphologies.

\end{enumerate}